\documentclass{aa}  

\usepackage{graphicx}
\usepackage{dblfloatfix}
\usepackage{caption}
\usepackage{nccmath}
\usepackage{txfonts}
\usepackage{natbib}
\bibpunct{(}{)}{;}{a}{}{,} 

\usepackage{hyperref}
\hypersetup{
colorlinks=true,
citecolor=blue
}

\begin{document} 

   \title{Cool circumgalactic gas of passive galaxies\\ from cosmological inflow}

   \author{Andrea Afruni\inst{1}\and Filippo Fraternali\inst{1}\and
           Gabriele Pezzulli\inst{2}
          }

   \institute{Kapteyn Astronomical Institute, University of Groningen,
              Landleven 12, 9747 AD Groningen, The Netherlands\\
              \email{afruni@astro.rug.nl}
         \and
             Department of Physics, ETH Z\"{u}rich, Wolfgang-Pauli-Strasse 27, 8093 Z\"                 {u}rich, Switzerland\\
             }
 
  \abstract
  {The circumgalactic medium (CGM) of galaxies consists of a multiphase gas with components at very different temperatures, from $10^ {4}$ K to $10^ {7}$ K. One of the greatest puzzle about this medium is the presence of a large amount of low-temperature ($T\sim10^4$ K) gas around quiescent early-type galaxies (ETGs).
Using semi-analytical parametric models, we describe the cool CGM around massive, low-redshift ETGs as the cosmological accretion of gas into their dark matter halos, resulting in an inflow of clouds from the external parts of the halos to the central galaxies. We compare our predictions with the observations of the COS-LRG collaboration. We find that inflow models can successfully reproduce the observed kinematics, the number of absorbers and the column densities of the cool gas. Our MCMC fit returns masses of the cool clouds of about $10^5\ \rm{M}_{\odot}$ and shows that they must evaporate during their journey due to hydrodynamic interactions with the hot gas.
We conclude that the cool gas present in the halos of ETGs likely cannot reach the central regions and feed the galaxy star formation, thus explaining why these passive objects are no longer forming stars.}


   \maketitle
%

\section{Introduction}\label{intro}
Characterizing the circumgalactic medium (CGM) and understanding its role in galaxy evolution is a key goal of current astrophysical models. Usually defined as the gas between the central galaxy and the surrounding intergalactic medium (IGM), this multiphase medium has been observed for decades in both emission and absorption from the X-ray \citep[e.g.][]{anderson13,li17,li18} to the UV-optical bands \citep{bordoloi11,keeney13,capra14,borthakur15} and is flowing both inward and outward galaxies, as suggested by numerical simulations \citep{ford14,suresh15,vandevoort12}. Despite the great theoretical and observational effort, however, the nature, origin and fate of this elusive medium are still much debated.\\
For a long time the most studied phase has been the hot CGM, also called the galaxy \textit{corona}, with temperatures of $T\sim10^{6-7}$ K, predicted decades ago by classical cosmological models (e.g. \citealt{white78,white91}, but see also \citealt{binney04}) as gas heated by shocks to the galaxy virial temperature. Although the observed luminosities are lower than predicted, hot gas in galaxy halos has been observed through X-ray observations both around early-type \citep{forman79,jones02,mathews03,bogdan11} and late-type galaxies \citep{dai12,li17}. The total amount of mass in this hot gas phase is however still uncertain \citep[e.g.][]{anderson11,bogdan13,li18}.\\
Recently, a number of investigations \citep{stocke06,rudie12,werk13,tumlinson13} have also focused on the cooler ($T<10^5$ K), low-density, ionized gas in the galaxy halos, originally discovered by \cite{boksenberg78} and \cite{bergeron86} and observed using UV and optical absorption lines in the spectra of background quasars. The presence of this cool gas seems ubiquitous around both external galaxies and the Milky Way, with high covering fractions \citep[e.g.][]{shull09,lehner11,werk14,borthakur16} and potentially large masses \citep{werk14,stocke13,stern16}. Understanding this cool medium is crucial in the study of the galaxy evolution and in solving the galaxy missing baryons problem \citep{mcgaugh}. So far our knowledge of this medium is largely limited and the general picture is still unclear, especially regarding its origin and dynamics. For star-forming galaxies, both transverse absorption-line studies and down-the-barrel observations of inflows \citep[e.g.][]{rubin12,lehner13,bouche13,borthakur15} and outflows \citep[e.g.][]{rubin14} suggest that the observed CGM dynamics is consistent with the recycling scenario \citep{ford14} in which the galaxy central enriched outflows are not able to escape from the potential well and then fall down again onto the galaxy. In this picture, the CGM is formed by a continuous cycle throughout the halo similar to the galactic fountain scenario proposed for smaller, galactic scales \citep{shapiro76,bregman80,fraternali06}.\\
The situation becomes however more complicated for passive early-type galaxies (ETGs). Cool ionized gas is in fact observed also in the CGM of passive massive galaxies, through observations of MgII \citep{gauthier09,huang16}, HI and other low-ion absorption lines \citep{thom12,chen18}. The presence of cool-enriched gas around quiescent galaxies remains a puzzle.
In fact, in the recycling scenario described above, a fundamental role is played by the central galaxy star formation, which causes the outflows thus starting the entire cycle. However, how can we explain the presence of cool gas in the halos of galaxies with little or no ongoing star formation? What mechanism can produce the cool gas and how is it prevented from fueling star formation in the central passive galaxy? The main goal of this work is to address these questions.\\
To this purpose, we use the observations of the COS-LRG collaboration \citep{chen18,zahedy19} who, using the Cosmic Origin Spectorgraph \citep[COS,][]{froning09} on the Hubble Space Telescope (HST), has observed the cool CGM in the halos of 16 luminous red galaxies (LRGs), massive elliptical galaxies at $z\sim 0.5$, where in principle cool gas is not expected to exist. One of the features of these observations is that the cool CGM is not homogeneous, but composed of different absorbing clouds, with different velocities, all bound to the central galaxies, a property of this medium that was already well known from previous absorption-line studies (e.g COS-Halos survey, see \citealt{werk13,werk14,tumlinson13}). Our purpose is to model the observed cool gas kinematics. Due to the absence of strong outflows in these galaxies, we assume that the cool CGM clouds are infalling toward the central galaxies.\\ 
As we will discuss in detail (Section~\ref{model}), the dynamics of the CGM -- and in particular the interactions between its different phases -- depends on a number of physical processes acting on both very large and very small physical scales. Ideally one would like to model the entire CGM with large-volume high-resolution hydrodynamical simulations. Unfortunately, this is still far out of the current computational capabilities: in fact, even the current "zoom-in" simulations \citep[e.g.][]{grand17,fattahi16}, which can trace a single galaxy halo, do not have enough resolution to resolve the small-scale CGM interactions \citep{armillotta17}.  We therefore adopt a semi-analytical modelling, keeping track of the gravitational force which dominates the dynamics on large scales, as well as taking into account prescriptions on hydrodynamical processes from small-scale high resolution simulations \citep[e.g.][]{heitsch09,mccourt15,armillotta16,gronnow18}.\\
This paper is organized as follows: in Section~\ref{obs} we briefly describe the COS-LRG observations, their characteristics and the main observables that we use as a comparison for our models; in Sections~\ref{model} and ~\ref{comparison} we describe the construction of the model and the comparison with observations; in Section~\ref{resdisc} we present our results while in Section~\ref{discussion} we discuss the implications of our findings for the properties, origin and fate of the cool CGM around massive ETGs; finally, Section~\ref{concl} contains the summary and the conclusions of the work.
\section{ETGs sample and observational constraints}\label{obs}
\begin{figure}[htbp]
   \includegraphics[width=9cm]{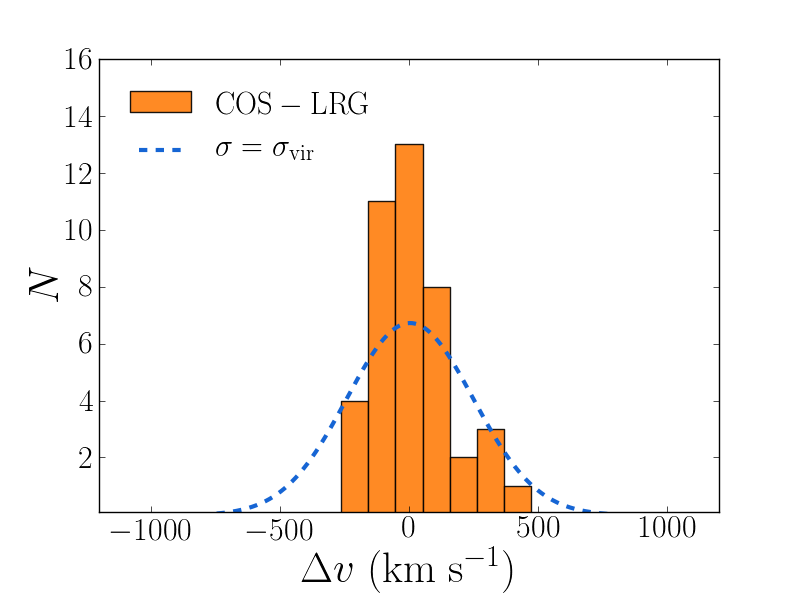}
   \caption{In orange, observed velocities of the cool CGM clouds relative to systemic velocities of the galaxies, obtained using the COS-LRG data for a sample of 16 massive passive galaxies (their properties are reported in Table~\ref{tab:galprop}). The light-blue dashed line is a gaussian with a dispersion equal to the typical 1D (line-of-sight) virial velocity dispersion value in the sample.}
              \label{fig:velobs}%
    \end{figure}

In this work we focus on the data of the COS-LRG program.
Here we briefly describe the main characteristics of these observations, while we refer, for further information and details, to the COS-LRG papers \citep{chen18,zahedy19}. This survey provides high-resolution kinematic data of the cool CGM around 16 massive early-type galaxies and it is therefore the most suitable for the purpose of this work, which aims to reproduce in particular the kinematics of the cool CGM clouds around passive galaxies. We assume that, due to the similarity of the objects in this sample, the CGM has the same behavior and properties in all these galaxy halos, modulo some scaling factors due to the slightly different halo masses and virial radii, which we will account for in our model. The main properties of the galaxies in the sample are reported in Table~\ref{tab:galprop}.\\
The CGM is characterized using HI and metal absorption lines in the COS and optical echelle spectra of UV-bright quasars with impact parameters (the projected distances between the quasars and the galaxies) $R<160$ kpc. 
A typical feature of these observations is the presence, for each line of sight, of different velocity components for the same ionic transition, characteristic of a medium that is not uniform, but composed of different clouds with different projected velocities, plausibly different intrinsic positions along the line of sight and possibly different physical properties (e.g. density or ionized fraction). This is studied in \cite{chen18} and \cite{zahedy19} fitting Voigt profiles and obtaining the velocity offsets of the single components from the galaxy systemic velocity $\Delta v$. In this paper we use in particular the analysis carried out by \cite{zahedy19}, who fitted the Voigt profiles imposing the same kinematic structure among HI, low-ionization and intermediate-ionization species (including MgII, SiII, SiIII and CIII). This choice is justified by the kinematic agreement found by \cite{chen18} in the absorption profiles of these different species. Each velocity component found by \cite{zahedy19} represents a single cloud: they find 42 clouds among all the galaxies (including 3 sightlines with no detections of absorption lines), with an average of 2.6 clouds per line of sight. In Figure~\ref{fig:velobs} we report all the 42 observed cool cloud velocities ($\Delta v$, relative to the galaxy systemic velocity) together in one single distribution. The two main observational constraints that we used for the comparison with our model results are then given by:
\begin{enumerate}
\item the cloud velocity distribution;
\item the total number of observed clouds.
\end{enumerate}
 Due to the high resolution of the COS spectra, these kinematic data are very accurate.
 {
\renewcommand{\arraystretch}{1.2}
 \begin{table*}[htbp]
\begin{center}
\caption[]{Properties of the galaxies in our sample.}\label{tab:galprop}
\begin{tabular}{*{9}{c}}
(1)&(2)&(3)&(4)&(5)&(6)&(7)&(8)&(9)\\
\hline  
\hline
Galaxy ID  & z  & $\log\ (M_{\ast}/\rm{M}_{\odot})$ & $\log\ (M_{\rm{vir}}/\rm{M}_{\odot})$ & $r_{\rm{vir}}$& $T_{\rm{vir}}$& $R$& $n_{\rm{cl}}$&$\log(N_{\rm{H}}/\rm{cm}^{-2})$  \\
               &     & & & (kpc) & ($10^6\ \rm{K})$ & (kpc)&  \\
\hline
J094632.40+512335.9 & 0.41 & 11.2 & 13.0 & 427 & 3.3 & 42 & 5 & 19.5\\
J140625.97+250923.2 & 0.40 & 11.1 & 12.9 & 397 & 2.8 & 47 & 5 & 19.2\\
J111132.33+554712.8 & 0.46 & 11.4 & 13.4 & 564 & 6.3 & 77 & 3 & 19.2\\
J080357.74+433309.9 & 0.25 & 11.1 & 12.9& 433& 2.6 &78&4& 18.6\\
J092554.18+400353.4 & 0.25 & 11.1 & 12.9& 435 &2.6&84&1& 19.9\\ 
J095000.86+483102.2 & 0.21 & 11.0 & 12.7& 381 &1.9& 94&4& 19.7\\
J112755.83+115438.3 & 0.42 & 11.2 & 12.9 & 423 & 3.3 & 99 & 2 & 17.7\\
J124307.36+353926.3 & 0.39 & 11.3 & 13.2 & 504 & 4.4 & 102 & - & -\\
J155047.70+400122.6 & 0.31 & 11.2 & 13.0& 452 &3.1& 107&4& 18.9\\
J024651.20-005914.1 & 0.41 & 11.4 & 13.4 & 581 &6.1 & 109 & 6 & 19.2\\
J135727.27+043603.3 & 0.33 & 11.3 & 13.2 & 522 &4.3 & 126 & 1 &19.0\\
J091027.70+101357.2 & 0.26 & 11.2  & 13.0 & 465 & 3.1&140&4& 18.5\\
J141307.39+091956.7 & 0.36 & 11.7 & 14.1 & 1028 & 17.6 & 149 & - & -\\
J155304.32+354853.9 & 0.47 & 11.0 & 12.6 & 326 &2.1 & 156 & 2 & 17.8\\
J125859.98+413128.2 & 0.28 & 11.6 & 13.8 & 855 & 10.7 & 159 & - & -\\
J124409.17+172111.9 & 0.56 & 11.5 & 13.5 & 623 & 8.9 & 160 & 3 & 18.2\\
\hline
\end{tabular}
\end{center}

\tablefoot{(1) Galaxy name (2) Galaxy redshift (3) Galaxy stellar mass (4) Galaxy virial mass, calculated using the relation of \cite{kravtsov18} as explained in Section~\ref{obs} (5) Virial radius, see Section~\ref{obs} (6) Virial temperature (equation~\ref{eq:Tvir}) (7) Projected distance between the line of sight and the galaxy (8) Number of observed clouds (9) Total hydrogen column density. All the data are taken from \cite{chen18,zahedy19}.}
   \end{table*}\\
   }
\cite{zahedy19} also performed a photo-ionization analysis for each kinematic component, using the code CLOUDY v13.03 \citep{ferland13}. This analysis allowed them to estimate the values of the total-hydrogen column densities $N_{\rm{H}}$ for each line of sight (reported in Table~\ref{tab:galprop}). Differently from the very accurate and reliable kinematic data, these values are however subject to various model-dependent uncertainties and therefore we do not use them as an observational constraint for our model fitting. We will discuss in Section~\ref{resdisc} the comparison between these data and our final results.\\
For the purpose of this work, we are interested in the virial masses and virial radii of the galaxies in the COS-LRG sample, which allow us to calculate the halo virial temperatures. To calculate these quantities we first obtained $M_{\rm{200}}$ using the stellar-to-halo mass relation (SHMR) of \cite{kravtsov18}. We note that at these high stellar masses the slope of the SHMR is uncertain and the galaxy halo masses could be higher using different prescriptions \citep[e.g.][]{moster13}. However, the relation from \cite{kravtsov18}, obtained with a recent stellar mass function based on improved photometry, is in agreement with estimates based on X-ray observations, weak lensing and satellite kinematics for objects with these high masses. Thus we consider this prescription the most reliable for the massive galaxies of our sample.
Once we obtained the values of $M_{200}$, we calculated $r_{\rm{200}}$ through the formula
\begin{ceqn}
\begin{equation}\label{eq:Rvir}
r_{\rm{200}}=\left(\frac{M_{\rm{200}}2G}{200H^2}\right)^{1/3},
\end{equation}
\end{ceqn}
where $H$ is the Hubble parameter (function of redshift, with $H_0=70\ \rm{km}\ \rm{s}^{-1}\ \rm{Mpc}^{-1}$) for the standard cosmological model ($\Omega_{\rm{m},0}=0.3$, $\Omega_{\Lambda,0}=0.7$). Using these values, we calculated the virial masses $M_{\rm{vir}}$ and radii $r_{\rm{vir}}$ extrapolating for each halo the dark matter (DM) profile (we assumed a Navarro Frenk White profile, see Section~\ref{grav}) to the radius with virial overdensity, which is $\Delta$ times the critical density of the universe, with $\Delta$ calculated using the prescription in \cite{bryan98} ($\Delta=119-141$, depending on the redshift of each galaxy).\\
Finally, the virial temperature is given by
\begin{ceqn}
\begin{equation}\label{eq:Tvir}
T_{\rm{vir}}=\frac{\mu m_{\rm{p}}GM_{\rm{vir}}}{2k_{\rm{B}}r_{\rm{vir}}}\ ,
\end{equation}
\end{ceqn}
where $\mu=0.6$ is the mean molecular weight for a hot, totally ionized gas \citep{sutherland93}, $m_{\rm{p}}$ is the proton mass and $k_{\rm{B}}$ is the Boltzmann constant. Our estimates of these virial quantities for each galaxy are reported in Table~\ref{tab:galprop}.\\
 As explained above, the kinematic distribution in Figure~\ref{fig:velobs} is our main constraint and our purpose is to create theoretical models that are able to reproduce a similar distribution, with a comparable shape and a comparable number of clouds. It is interesting to note from this observed distribution that the clouds show an unexpected narrow distribution (small $\Delta v$). Indeed, the line-of-sight virial velocity dispersion for a galaxy halo with a mass equal to the mean value of the virial masses of our sample ($M_{\rm{vir}}=10^{13.3}\ \rm{M}_{\odot}$) is equal to 245 km $\rm{s}^{-1}$, while the velocity dispersion of the distribution in Figure~\ref{fig:velobs} is equal to only 147 km $\rm{s}^{-1}$. If the cloud motion were related only to the dynamical mass of the galaxies we would expect the clouds to have a dispersion comparable to the virial one. The small observed velocity dispersion, instead, suggests that there should be some mechanism that slows down these absorbers during their motion throughout the halos \citep[see also][]{huang16} and this is a fundamental observational feature that our models aim to reproduce.\\
It is particularly important to compare our model results not only with the cloud velocities, but also with the number of observed clouds, because with this additional constraint we have more information on the total mass accretion rate and cloud masses that are needed to successfully reproduce the observations (see Section~\ref{model}), breaking the degeneracy between different cloud infall models.\\
One limitation of these data is that they only contain observations at relatively small projected radii: we have information only for $R<160$ kpc, which is much smaller than the virial radii of these massive galaxies. The average virial radius of these halos is indeed 526 kpc. We discuss the comparison between our results and other observations at larger distances later in  Section~\ref{discussion}.
\section{The model}\label{model}
In this Section we describe how we built our dynamical models, starting from the idea that the cool CGM is composed of different clouds (see Sections~\ref{intro} and \ref{obs}), which originate from the cosmological accretion of gas into the galaxy halos and are infalling toward the central galaxies. We first describe, in Sections~\ref{grav} and \ref{hydro}, how we calculated the infall velocities of the clouds for each of the sixteen galaxies of our sample; then in Section~\ref{accrrate} we focus on the total accretion rate of cool gas inside our model halos.
\subsection{Purely ballistic model}\label{grav}
The main force that drives the motion of the cool clouds is the gravitational force of the dark matter halo. Gravity forces the cool medium to accrete toward the central galaxies. If we considered only this force, the equation of motion would be:
\begin{ceqn}
\begin{equation}\label{eq:grav}
\frac{\rm{d} \textit{v}_{\rm{fall}}}{\rm{d}\textit{r}}=\frac{1}{v_{\rm{fall}}(r)}\frac{GM(r)}{r^2},
\end{equation}
\end{ceqn}
where $M(r)$ is the DM mass within the radius $r$. Throughout this work, we consider the velocity $v$ with a positive sign if it is pointing toward the center.\\
We have assumed in our treatment a Navarro Frenk White (NFW) profile \citep{nfw95}, described by
\begin{ceqn}
\begin{equation}\label{eq:NFW}
M(r) = 4\pi \rho_{0} r_{\rm{s}}^3 \displaystyle\left[\ln(1+r/r_{\rm{s}})-\frac{r/r_{\rm{s}}}{1+r/r_{\rm{s}}}\right]\ ,
\end{equation}
\end{ceqn}
where
\begin{ceqn}
\begin{equation}\label{eq:rozero}
\rho_0 = \frac{M_{\rm{vir}}}{4\pi r_{s}^3 \displaystyle\left[\ln(1+r_{\rm{vir}}/r_{\rm{s}})-\frac{r_{\rm{vir}}/r_{\rm{s}}}{1+r_{\rm{vir}}/r_{\rm{s}}}\right]}\ ,
\end{equation}
\end{ceqn}
is the central density, $M_{\rm{vir}}$ and $r_{\rm{vir}}$ are the virial masses and radii reported in Table~\ref{tab:galprop} and $r_{\rm{s}}=r_{\rm{vir}}/c$ is the scale radius, where $c$ is the concentration calculated for each halo following \cite{dutton14}.
   \begin{figure}[htbp]
   \includegraphics[width=9cm]{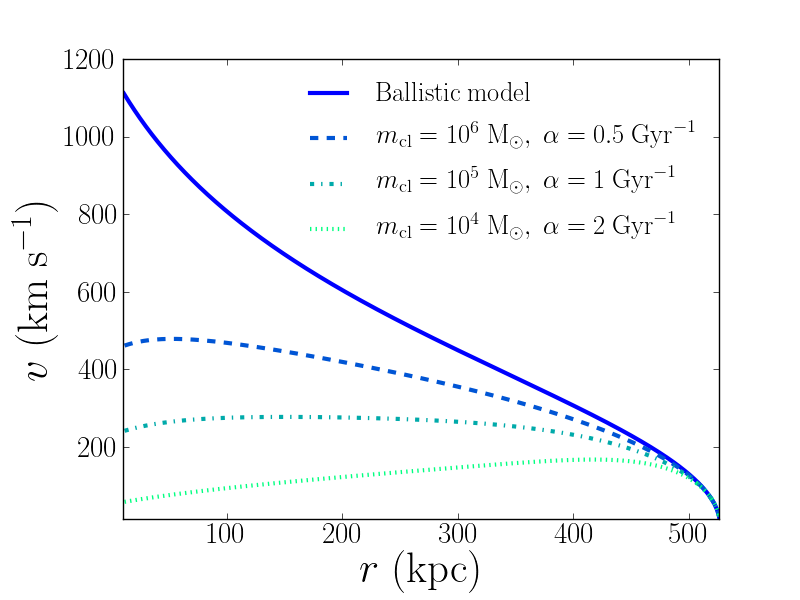}
   \caption{Velocity profiles of the cool CGM clouds obtained using the ballistic model (equation~\ref{eq:grav}, blue line) and the final model that includes all the hydrodynamical interactions (equation~\ref{eq:motion}), with three different choices of $m_{\rm{cl,start}}$ and $\alpha$.}\label{fig:velprofiles}%
    \end{figure}\\
By solving equation~\eqref{eq:grav} for a characteristic galaxy of our sample ($M_{\rm{vir}}=10^{13.3}\ \rm{M}_{\odot}$ and $r_{\rm{vir}}=526$ kpc, obtained averaging the values of the 16 galaxies in Table~\ref{tab:galprop}), assuming that the clouds start to fall with $v=0\ \rm{km}\ \rm{s}^{-1}$ from the virial radius, we can obtain the velocity profile shown as a blue line in Figure~\ref{fig:velprofiles}. Considering only the gravitational force would lead all the clouds to accrete onto the central galaxies with velocities higher than $1000\ \rm{km}\ \rm{s}^{-1}$ and this prediction is totally inconsistent with the low velocities observed by COS-LRG (see Figure~\ref{fig:velobs}).
We conclude that a simple ballistic infall is excluded by the observations and we move to more accurate models. As explained in Section~\ref{intro}, the CGM is a multiphase gas and the cool gas clouds must be surrounded by a hotter medium (corona) which influences their motion. Therefore, to describe the kinematics of the clouds we need, in addition to the gravitational force, to introduce in our model also hydrodynamical effects.
\subsection{Hydrodynamical effects}\label{hydro}
\begin{figure}[htbp]
   \includegraphics[width=9cm]{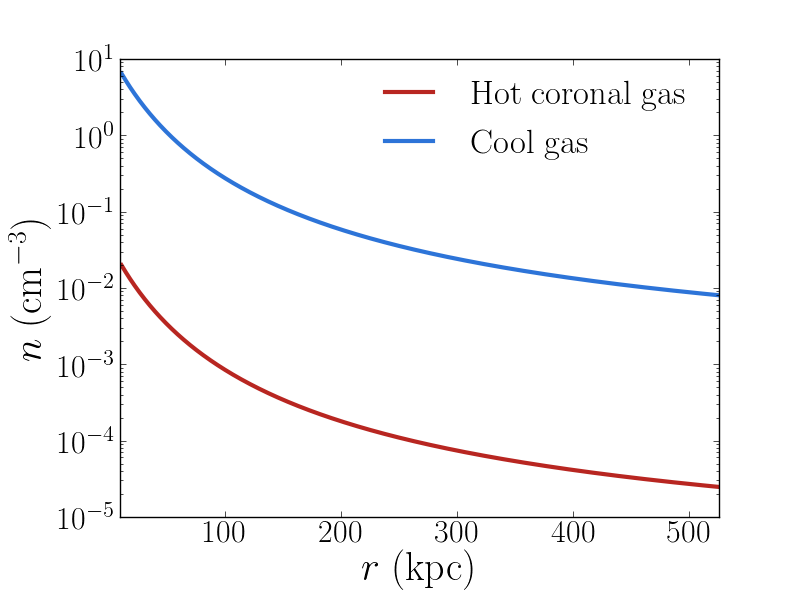}
   \caption{In red, profile of the intrinsic density of the hot coronal gas (from equation~\ref{eq:corona}) for a galaxy with $M_{\rm{vir}}=10^{13.3}\ \rm{M}_{\odot}$, $r_{\rm{vir}}=526$ kpc, $T =  T_{\rm{vir}}=6.5 \times 10^6$ K (average of the galaxy halos in our sample) and normalization set as explained in Section \ref{hydro}. In blue, profile of the intrinsic average density expected for the cool circumgalactic medium if in pressure equilibrium with the corona, obtained through equation~\eqref{eq:eqpress}.}
              \label{fig:presseq}%
    \end{figure}

To simulate the interactions between the cool clouds and the hot coronal gas, we decided to use an analytical parametrization of the hydrodynamical effects that take place in the CGM. 
Using an analytic treatment has several advantages, including computational speed, flexibility and physical insight. Our treatment matches however indications from high-resolution hydrodynamical simulations (see below).\\
First, we defined, for each galaxy of the sample, the corona as a gas at the galaxy virial temperature (\ref{eq:Tvir}), in hydrostatic equilibrium with the dark matter halo (\ref{eq:NFW}), which leads to the electron density profile
\begin{ceqn}
\begin{equation}\label{eq:corona}
n_{\rm{e,cor}}(r)=n_{\rm{e,0}}\exp \displaystyle \left[-\frac{\mu_{\rm{cor}} m_{\rm{p}}}{k_{\rm{B}}T_{\rm{vir}}}(\Phi(r)-\Phi_0)\right]\ ,
\end{equation}
\end{ceqn}
where
\begin{ceqn}
\begin{equation}\label{eq:potNFW}
\Phi(r)=4\pi G \rho_{0}r_{\rm{s}}^2 \left[ \ln \left( \frac{1+r}{r_{\rm{s}}} \right)\frac{r_{\rm{s}}}{r} \right]\ 
\end{equation}
\end{ceqn}
is the NFW potential. Here, $\rho_0$ and $r_{\rm{s}}$ are the same as in equation~\eqref{eq:NFW}, $k_{\rm{B}}$ is the Boltzmann constant, $m_{\rm{p}}$ is the proton mass, $\mu_{\rm{cor}}=0.6$ is the mean molecular weight and $n_{\rm{e,0}}$ is the normalization of the gas profile. We fixed $n_{\rm{e,0}}$ by requiring that the total coronal mass is equal to 20\% of the total baryonic mass within the galaxy halo (assuming $M_{\rm{bar}}=0.158 M_{\rm{vir}}$, where 0.158 is the baryon fraction as in \citealt{planck18}). The 20\% value is justified by observational estimates \citep[e.g.][]{anderson11,bogdan13}, although the exact amount of hot gas present in the galaxy halos is still debated.\\
We assumed then that the cool CGM clouds are pressure confined by the hot gas, using the following formula
\begin{ceqn}
\begin{equation}\label{eq:eqpress}
n_{\rm{cl}}T_{\rm{cl}}=n_{\rm{cor}}T_{\rm{vir}}
\end{equation}
\end{ceqn}
where $T_{\rm{cl}}=2 \times 10^4$ K is the characteristic temperature of the cool absorbers \citep{zahedy19} and $n_{\rm{cor}}=2.1n_{\rm{e,cor}}$ is the total particle number density in the corona.
Figure~\ref{fig:presseq} shows as an example the density profiles of the hot and cool mediums for an average galaxy with the same properties as the one used in Section~\ref{grav}. The coronal density is low at large distances from the central galaxy, while it increases in the central regions of the halo. Due to the pressure equilibrium (\ref{eq:eqpress}), also the densities of the cool clouds, represented by the blue line in Figure~\ref{fig:presseq}, will increase at lower distances from the central galaxy.\\
The first effect of the hot gas on the cloud motion is the deceleration due to the drag force, that we can write \citep[e.g.][]{marinacci11}
\begin{ceqn}
\begin{equation}\label{eq:drag}
\dot{v}_{\rm{drag}}=\frac{\pi r^2_{\rm{cl}}\rho_{\rm{cor}}v^2}{m_{\rm{cl}}},
\end{equation}
\end{ceqn}
where $\rho_{\rm{cor}}=\mu_{\rm{cor}}\ m_{\rm{p}} n_{\rm{cor}}$, $v$ is the cloud velocity, $m_{\rm{cl}}$ is the cloud mass and $r_{\rm{cl}}$ is the cloud radius, given by
\begin{ceqn}
\begin{equation}\label{eq:radii}
r_{\rm{cl}}=\left(\frac{3m_{\rm{cl}}}{4\pi \rho_{\rm{cl}}} \right)^{1/3}\ ,
\end{equation}
\end{ceqn}
where $\rho_{\rm{cl}}=\mu_{\rm{cl}} m_{\rm{p}} n_{\rm{cl}}$ is the density of the cool medium and $\mu_{\rm{cl}}=0.6$ is the mean molecular weight for a cool gas in photo-ionization equilibrium. We note that a slightly higher value $\mu_{\rm{cl}}=0.67$ \citep{sutherland93} would be appropriate if the cool gas is in collisional ionization equilibrium, which is however uncertain and would have a negligible impact on our results.\\
Since the density of the cool gas increases with the decrease of the distance from the galaxy, the clouds become smaller during their motion toward the galaxies.
Adding the drag force term to equation~\eqref{eq:grav} decelerates the clouds during their infall, with a dependence on the cloud mass.
Less massive clouds are affected by a stronger drag force and exhibit a higher deceleration.  As there are no observational constraints on the mass of the clouds, we let it vary as a free parameter in our model.
   \\
With the inclusion of the drag force in the treatment, the results are more accurate and similar to the real behavior of the clouds.
However, this simple model still does not take into account the hydrodynamical instabilities that are expected to take place in the system. High-resolution simulations \citep[e.g.][]{armillotta17} show how clouds moving through a hot-gas environment lose mass with time, mostly because of instabilities that strip cool gas from the cloud and mix it with the hotter medium. The result of these interactions is the evaporation of the cool cloud in the hot gas. These mass losses obviously affect the cloud velocities and can significantly change the output of the models.
For this reason, we introduced in our modeling a mass loss term, described by
\begin{ceqn}
\begin{equation}\label{eq:massloss}
\frac{\rm{d} \textit{m}_{\rm{cl}}}{\rm{d}\textit{r}}=-\frac{\alpha m_{\rm{cl}}}{v}\ .
\end{equation}
\end{ceqn}
Equation~\eqref{eq:massloss} approximates the hydrodynamical instabilities, by making the clouds lose mass during their fall at a certain constant evaporation rate $\alpha$ \citep[see][]{fraternali08,marinacci10}. As for the mass, the exact value of $\alpha$ is unknown and we thus included it in the model as a second free parameter. Since, with the inclusion of equation~\eqref{eq:massloss}, the mass of the clouds is a function of the intrinsic distance, the free parameter that we vary is the mass of the clouds at the starting radius of their infall motion, $m_{\rm{cl,start}}$.\\
The inclusion of all the effects described in equations~\eqref{eq:grav}, \eqref{eq:drag} and \eqref{eq:massloss} leads to the following final equation of motion:
\begin{ceqn}
\begin{equation}\label{eq:motion}
\frac{\rm{d}\textit{v}}{\rm{d}\textit{r}}=\frac{1}{v(r)}\frac{GM(r)}{r^2}-\frac{\pi r^2_{\rm{cl}}(r)\rho_{\rm{cor}}(r)v(r)}{m_{\rm{cl}}(r)}\ .
\end{equation}
\end{ceqn} 
In order to solve equation~\eqref{eq:motion}, the starting velocity $v_{\rm{start}}$ of the clouds is needed as an initial condition. We adopted $v_{\rm{start}}=0\ \rm{km}\ \rm{s}^{-1}$, since we have found that to reproduce the observed velocity distribution the clouds must have very low initial velocities (see Appendix~\ref{invel} for details).
We also fixed the starting radius of the cloud infall motion at the virial radius of the galaxies (Table~\ref{tab:galprop}). This choice is justified from the observations of the cool circumgalactic medium extending till distances comparable to the virial radii of these galaxies \citep{zhu14,huang16} and from our assumption that these clouds are inflowing from the external parts of the halos toward the center. We have also explored models with different starting radii for the clouds (Section~\ref{discussion}).\\
Solving equation~\eqref{eq:motion} we can find the velocities of the clouds with respect to their distance $r$ from the galaxy center, taking into account both the gravitational force and the hydrodynamical effects.
The different curves in Figure~\ref{fig:velprofiles} show how the velocity profile varies with different choices of the parameters, using the same average galaxy used in Section~\ref{grav} for the ballistic motion and the coronal model in Figure~\ref{fig:presseq}. After being initially accelerated by the gravitational force the clouds will be decelerated by the interactions with the corona. Models with lower cloud masses and higher evaporation rates have, particularly in the central regions, clouds with lower velocities, due to the higher efficiency of the hydrodynamical interactions. It is therefore crucial to explore the parameter space to understand which choice of the parameters leads to a velocity profile consistent with what we observe.
\subsection{Accretion rate}\label{accrrate}
In the model outlined above the clouds start from the virial radius and then are attracted by the gravitational force toward the center of the halos. The last step in creating the model is to define the total cool gas accretion rate at the the virial radius.\\
To this purpose, we assumed that the CGM cool clouds are coming from the cosmological gas accretion. Cosmological models in fact predict that the DM halos are growing in time by the accretion of external matter. We can estimate the quantity of matter accreted per unit time by using the prescription in \cite{fakhouri10}, who proposed a fit for the mass growth rate of the DM halos based on the results of the Millennium simulations \citep{springel05,boylan09}. Multiplying this prescription by the baryonic fraction, we obtain the rate of baryonic matter accreting in the halos:
\begin{ceqn}
\begin{equation}\label{eq:cosmaccr}
\begin{aligned}
\dot{M}_{\rm{cosm}} =47.6\ h^{-1}\rm{M}_{\odot}\rm{yr}^{-1}\left(\frac{M_{\rm{vir}}}{10^{12}h^{-1}\rm{M}_{\odot}} \right)^{1.1}(1+1.11z)\\  \times \sqrt{\Omega_{\rm{m},0}(1+z)^3+\Omega_{\Lambda,0}} \times f_{\rm{b}}
\end{aligned}
\end{equation}
\end{ceqn}
where $h=0.7$, $\Omega_{\rm{m},0}=0.3$, $\Omega_{\Lambda,0}=0.7$ and $f_{\rm{b}}= 0.158$ is the baryon fraction by \cite{planck18}.\\
We related the accretion rate to the cosmological accretion, using the formula
\begin{ceqn}
\begin{equation}\label{eq:normalization}
\dot{M}_{\rm{accr}}(r_{\rm{vir}})=f_{\rm{accr}}\dot{M}_{\rm{cosm}}
\end{equation}
\end{ceqn}
where $f_{\rm{accr}}$ is the fourth and last free parameter that we need in our analysis. If the value of $f_{\rm{accr}}$ is equal to 1, all the gas predicted by cosmological models to be accreted into the galaxy halo is what we observe as cool circumgalactic medium.\\
Once we have defined the value of the accretion rate at the virial radius, we can estimate the mass flux rate as a function of the intrinsic radius, using the formula:
\begin{ceqn}
\begin{equation}\label{eq:continuity}
\dot{M}_{\rm{accr}}(r)=4\pi \rho_{\rm{cool}}(r) v(r)r^2=\dot{M}_{\rm{accr}}(r_{\rm{vir}})f_{\rm{mass}}(r)
\end{equation}
\end{ceqn}
where $\rho_{\rm{cool}}(r)$ is the volume averaged mass density of the cool gas and $f_{\rm{mass}}=m_{\rm{cl}}(r)/m_{\rm{cl,start}}$ is a term that takes into account the mass losses of the clouds at every radius. In this way, the mass flux decreases at smaller distances from the center, consistently with the evaporation of the clouds in the hot gas. For models with $\alpha=0$, which means no evaporation of the clouds, the mass flux is constant with radius.\\
Inverting equation~\eqref{eq:continuity} we obtained the value of $\rho_{\rm{cool}}(r)$, which will be useful in the next Section in order to compare the results with the observations.
\section{Comparison with the observations}\label{comparison}
In this Section we explain how we obtained, starting from the intrinsic quantities calculated in Section~\ref{model}, the results that are directly comparable to the observations. To this purpose, we created random distributions of cool clouds over the whole halo, each of them with the properties defined by the model, and we 'observed' them through synthetic observations.\\
The quantity $\rho_{\rm{cool}}(r)$ describes the total mass density of the cool gas per unit of volume. If we divide this density by the mass of the clouds, which is a function of the intrinsic radius, described by equation \eqref{eq:eqpress}, we obtain the number of clouds per unit of volume. The integral of this quantity over the whole volume of the halo is equal to the total number of cool absorbers expected in the system:
\begin{ceqn}
\begin{equation}\label{eq:totnumber}
N_{\rm{cl}}=4\pi \int^{r_{\rm{vir}}}_{0}\frac{\rho_{\rm{tot}}(r)r^2}{m_{\rm{cl}}(r)}dr\ .
\end{equation}
\end{ceqn}
We created then a distribution with a total number of objects given by equation~\eqref{eq:totnumber}. The clouds are not distributed uniformly in the galaxy halo, as their probability to be at a certain radius $r$ is predicted by the argument of the integral in equation~\eqref{eq:totnumber}. Therefore, we populated the halos using a probability density function (PDF) given by:
\begin{ceqn}
\begin{equation}\label{eq:pdf}
\frac{4\pi \rho_{\rm{tot}}(r)r^2}{m_{\rm{cl}}(r)N_{\rm{cl}}}
\end{equation}
\end{ceqn}
We have then $N_{\rm{cl}}$ objects, each of them located at a certain radius $r$ and with all the properties given by the results of equations~\eqref{eq:eqpress}, \eqref{eq:radii}, \eqref{eq:massloss} and \eqref{eq:motion}. We associated then to every cloud two other coordinates $\theta$ and $\phi$, taken in order to have random distribution of clouds over a sphere (see Appendix~\ref{geometry} for the detailed 3-D geometry).
   \begin{figure}[ht!]
   \includegraphics[width=1\linewidth]{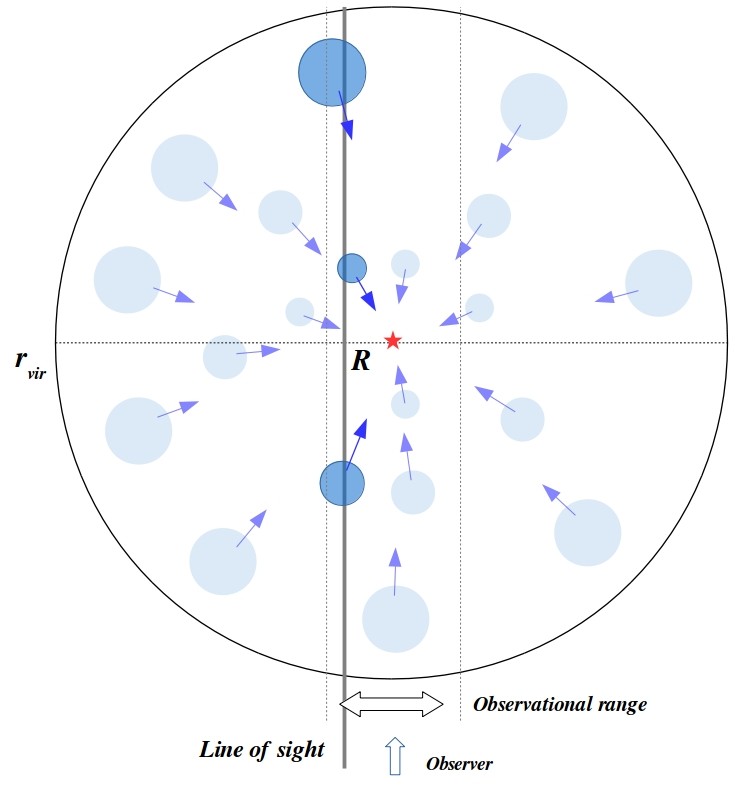}
   \caption{Cartoon 2-D representation of the synthetic observations performed in our model. The clouds (blue circles) are distributed throughout the halo (the sphere with radius $r_{\rm{vir}}$), all accreting towards the central galaxy (the central red star) with a radial velocity described by equation~\eqref{eq:motion}. The vertical gray line represents the line of sight, consistent with the projected distance $R$ of the galaxy from the background quasar, while the area included between the two vertical dashed lines represents the range spanned by the COS-LRG observations. The observer is located at the bottom side of the panel.}
              \label{fig:clvel}%
    \end{figure}\\
This setup was created for each one of the galaxies reported in Table~\ref{tab:galprop}. Figure~\ref{fig:clvel} is a cartoon representation of the cloud distribution, flattened in 2 dimensions: the clouds are randomly distributed throughout the halo, each one of them with the properties explained above. The last step was to create synthetic observations: we selected the 'observed' clouds tracing a line-of-sight through the galaxy halos, as explained in detail in Appendix~\ref{geometry} and shown in Figure~\ref{fig:clvel}. 
It is important to note that at the same projected distance we can intercept clouds at very different intrinsic distances from the central galaxy, potentially till the virial radius of the halo.
   \begin{figure*}[ht!]
   \centering
   \includegraphics[clip, trim={0 0 0cm 0}, width=14.9cm]{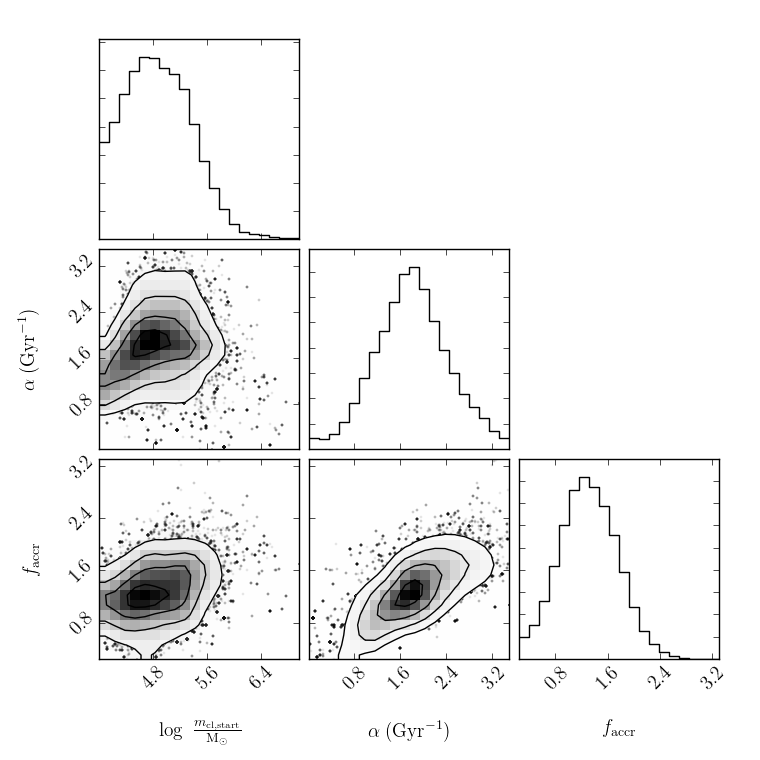}
   \caption{Corner plot with the MCMC results, representing the one and two dimensional projections of the posterior probabilities for the three free parameters of our models.}
              \label{fig:MCMC}%
    \end{figure*}\\
After selecting the clouds, we calculated their velocity projections along the line of sight (Appendix~\ref{geometry}). Figure~\ref{fig:clvel} shows the range of the projected distances between the galaxies and the quasars in the COS-LRG observations. As already mentioned in Section~\ref{obs}, these impact parameters are all much smaller than the typical virial radii of our massive galaxies. Therefore, the observations span only a small central area  of the halos. As a result, most of the clouds that we observe have a direction of the motion that is almost parallel to the line-of-sight and the projection corrections of the velocities are relatively small.
We repeated this procedure for all the galaxies and finally we put together all the observed velocities in a single line-of-sight velocity distribution.\\
The modeling described above allowed us to find the choice of parameters for our models that best reproduces the observations.
To this purpose, we performed a Markov Chain Monte Carlo (MCMC) analysis over the parameter space defined by $m_{\rm{cl,start}}$, $\alpha$ and $f_{\rm{accr}}$. We now describe the likelihood of the models, needed to perform the MCMC.
This is obtained using the comparisons between our models results and the two observables outlined in Section~\ref{obs}: the line of sight velocity distribution and the number of clouds. The total likelihood is the product of the likelihoods associated to these two observational constraints. In particular, we used the logarithm of the likelihood
\begin{ceqn}
\begin{equation}\label{eq:lnliketot}
\ln\mathcal{L}_{\rm{tot}}=\ln\mathcal{L}_1+\ln\mathcal{L}_2
\end{equation}
\end{ceqn}
where $\mathcal{L}_1$ and $\mathcal{L}_2$ are the two different likelihoods.
   \begin{figure*}[ht!]
   \includegraphics[width=.5\linewidth]{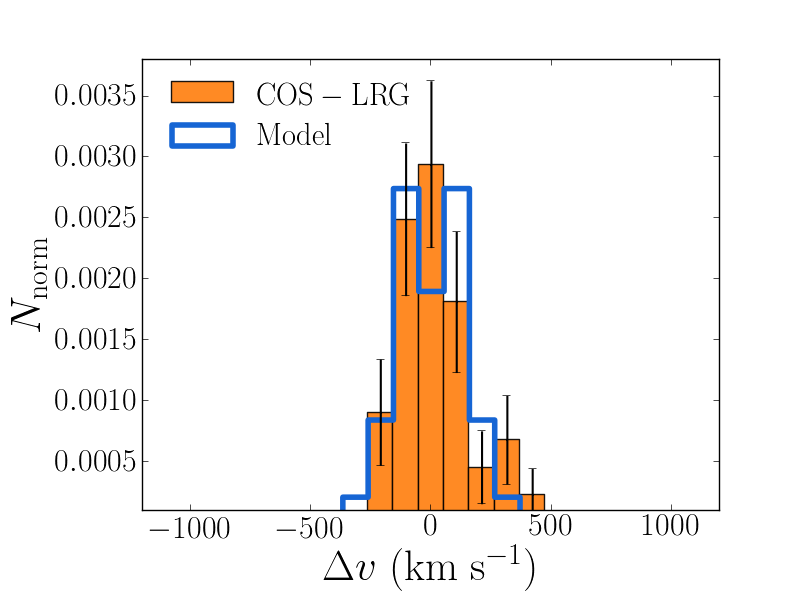}
   \includegraphics[width=.5\linewidth]{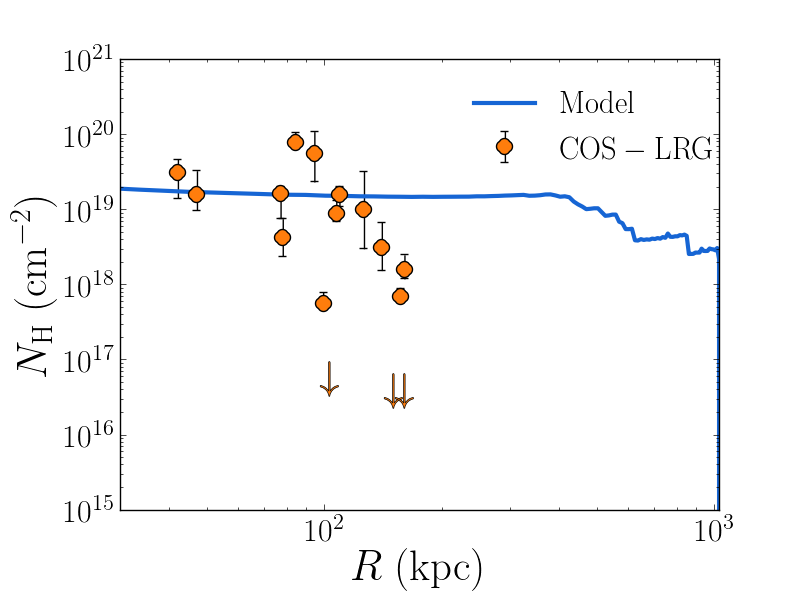}
   \caption{Comparison between the observations (orange bars and circles) and our best model results (blue lines). Left panel: normalized line-of-sight velocity distribution, where the errors in the observations are obtained with the bootstrapping method; right panel: total hydrogen column density. The downward arrows represent the upper limits of the non-detections.}
              \label{fig:comparisons}%
    \end{figure*}\\
To compare the two velocity distributions we have to assume for both of them the same number of velocity bins $n_{\rm{bin}}$ in which the velocity range is divided. We considered a velocity range spanning from $-1000$ to $1000\ \rm{km}\ \rm{s}^{-1}$, to include all the possible velocities predicted by our models, and we divided it in 20 different bins. However, we found that slightly different choices of the number of bins do not affect the final results.
The comparison between the two velocity distributions is then given by the reduced chi squared:
\begin{ceqn}
\begin{equation}\label{eq:lnlike1}
\ln\mathcal{L}_1=-\frac{1}{2(n_{\rm{bin}}-3)}\sum\limits_{b_i}\frac{\left|d_i-m_i\right|^2}{\sigma_i^2}\ ,
\end{equation}
\end{ceqn}
where $n_{\rm{bin}}=20$, $n_{\rm{bin}}-3$ is the number of degrees of freedom, $d_i$ and $m_i$ are respectively the values of the observed and model distributions in the velocity bins $b_i$ and $\sigma_i$ are the errors on the observed values, calculated using the bootstrapping method\footnote{We created 1000 randomizations of the observed velocity distribution: for each distribution, the number of objects is the same of the original one and its elements are randomly taken from the original distribution with the possibility of replacement. Then we divided all the distributions in the same number of velocity bins and we calculated for each of them the standard deviation $\sigma_i$.}.\\
Finally, the second constraint is given by the comparison between the number of clouds 'observed' in our synthetic models ($n_{\rm{mod}}$) and the total number of clouds observed by COS-LRG ($n_{\rm{obs}}=42$, see Section~\ref{obs}), using
\begin{ceqn}
\begin{equation}\label{eq:lnlike2}
\ln\mathcal{L}_2=-\frac{1}{2}\left(\frac{\left|n_{\rm{obs}}-n_{\rm{mod}} \right|}{\sqrt{n_{\rm{obs}}}} \right)^2
\end{equation}
\end{ceqn}
where $\sqrt{n_{\rm{obs}}}$ is the standard deviation of the Poisson distribution with a mean value equal to $n_{\rm{obs}}$.
The priors used in this analysis are flat for all the parameters but $f_{\rm{accr}}$, which sets the value of the total mass accretion rate of cool gas at the virial radius. In particular we employed a gaussian prior centered on 1 and with a dispersion of 0.5, in order to have the accretion rate consistent with the total baryonic accretion rate predicted by cosmology (equation~\ref{eq:cosmaccr}). Negative values of $f_{\rm{accr}}$ are excluded. Regarding the other two parameters, the logarithm of the cloud mass (in solar masses) is allowed to vary uniformly in the range from 4 to 7, while the evaporation rate $\alpha$ can vary uniformly from 0 to 3.5 $\rm{Gyr}^{-1}$. A negative $\alpha$ would in fact mean that the clouds are gaining cool mass and we excluded this scenario, because condensation is highly unlikely at these large temperatures: the cooling time of the hot gas is indeed too high for it to cool down and increase the amount of cold gas in the system. This is also supported by hydrodynamical simulations \citep{armillotta16}.  
\section{Results}\label{resdisc}
In this Section we report the results of the MCMC analysis that we used to compare our results to the observations and to find the best models that describe the CGM of massive ETGs. Then we discuss the physical scenario arising from our models.
\subsection{MCMC results}\label{mcmcres}
The results of the MCMC study are reported in Figure~\ref{fig:MCMC} and Table~\ref{tab:mcmcres}. Figure~\ref{fig:MCMC} shows the corner plot where both the one and two dimensional projections of the posterior probabilities of the three parameters are shown, while  in Table~\ref{tab:mcmcres} we show the 32nd, 50th (the median value) and 68th percentiles of the parameter one dimensional probabilities. Models with parameters in the ranges reported in Table~\ref{tab:mcmcres} are the best in maximizing the total likelihood $\mathcal{L}_{\rm{tot}}$ and therefore the best in reproducing the COS-LRG observations.\\
{
\renewcommand{\arraystretch}{1.3}
\begin{table}[htbp]
\begin{center}
\caption[]{Results of the MCMC analysis.}\label{tab:mcmcres}
\begin{tabular}{*{4}{c}}
\hline  
\hline
 Parameter & 32nd & 50th & 68th\\
\hline 
$\log\ (m_{\rm{cl,start}}/\rm{M}_{\odot})$ & 4.62 & 4.85 &5.1\\
$\alpha\ (\rm{Gyr}^{-1})$ & 1.54 & 1.8 & 2.05\\
$f_{\rm{accr}}$ & 1.11 & 1.29 & 1.5\\
\hline
\end{tabular}
\end{center}
\tablefoot{Best values (50th percentile), 32nd and 68th percentiles for the three free parameters.}
   \end{table}
}
   \begin{figure*}[ht!]
   \includegraphics[width=.5\linewidth]{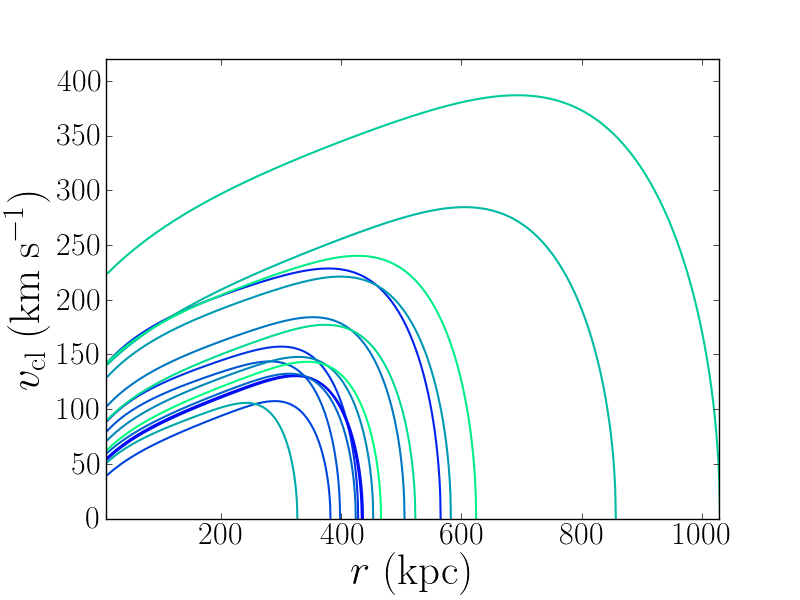}
   \includegraphics[width=.5\linewidth]{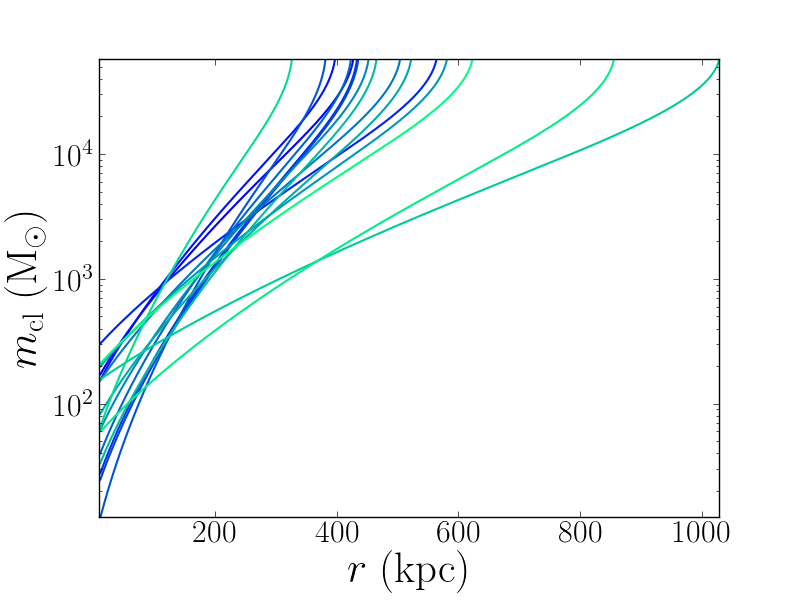}
   \includegraphics[width=.5\linewidth]{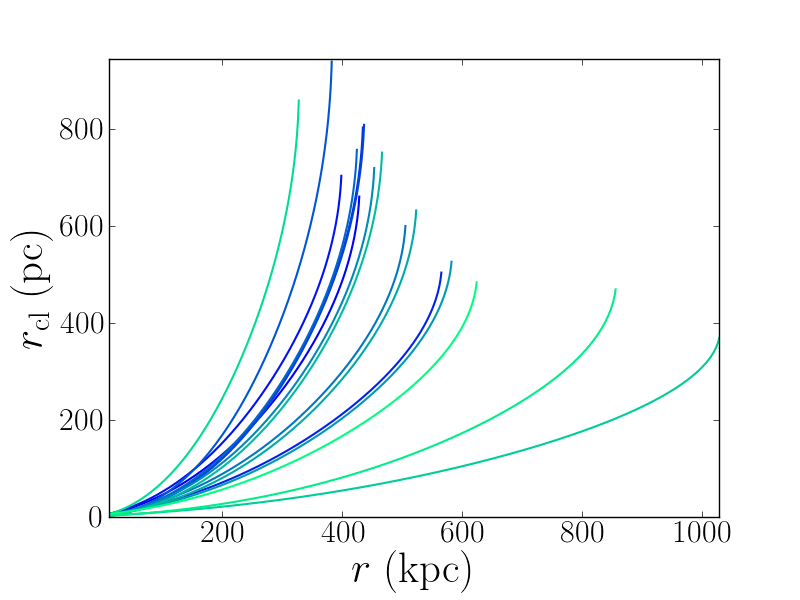}
   \includegraphics[width=.5\linewidth]{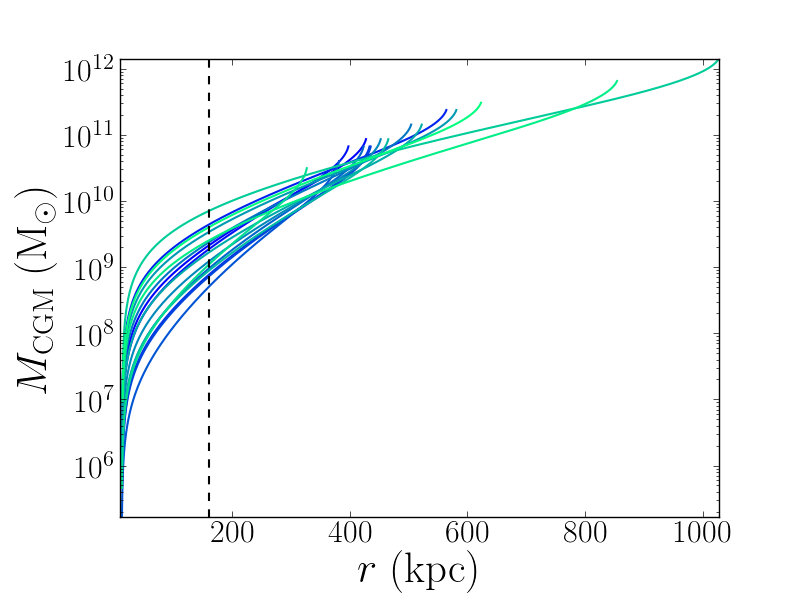}
   \caption{Properties of the cool CGM clouds described by our best model. All the profiles are plotted with respect to the intrinsic distance from the central galaxies and the different colored lines represent the result for the sixteen different galaxies of the sample. Top left: velocity profiles of the cool clouds; top right and bottom left: respectively masses and radii of the cool clouds; bottom right: cool CGM cumulative mass profile, with the vertical dashed line representing the distance of 160 kpc.}
              \label{fig:resprof}%
    \end{figure*}\\
In the left and right panels of Figure~\ref{fig:comparisons}  we show the outputs of these models, respectively the line of sight velocity distribution and the total hydrogen column densities, with their comparison with the observations. The errors in the observed distribution are calculated using the bootstrapping method explained in Section~\ref{comparison}. The model velocity distribution is dependent on the random populations of clouds, as explained in Section~\ref{comparison}. Therefore, to have a more robust comparison between our model output and the observations, the model distribution in the left panel of Figure~\ref{fig:comparisons} is the result of the average of 2000 models with the values of the parameters included within the 32nd and the 68th percentiles reported in Table~\ref{tab:mcmcres}, weighted with the value of the posterior probability of each model. The model and observed velocity distributions are both normalized to have a subtended area equal to 1. The kinematics of the model clouds, fitted by the MCMC analysis, is in good agreement with the observed one. The consistency between the two distributions is also evident from the comparison between the Cumulative Distribution Functions (CDFs) in Figure~\ref{fig:CDF}. The result of a Kolmogorov-Smirnov test between the two samples confirms that they are likely drawn from the same distribution. The averaged number of clouds of the 2000 randomizations used to create the plot in the left panel of Figure~\ref{fig:comparisons} is $43\pm10$ (the error is given by the weighted standard deviation), consistent with the observed one of 42.\\
The hydrogen column density of our model (blue curve) in the right panel of Figure~\ref{fig:comparisons} is instead obtained converting the density of clouds $\rho_{\rm{tot}}$ (see Section~\ref{accrrate}) into a numeric hydrogen density and projecting it along the line of sight. More precisely, the model curve in Figure~\ref{fig:comparisons} is the average of the sixteen column densities calculated for each galaxy. This profile is not dependent on the random populations of clouds and therefore we show here the result of a single model with the parameters given by the median values reported in Table~\ref{tab:mcmcres}. The orange circles represent instead the column density observations by \cite{zahedy19}, also reported in Table~\ref{tab:galprop}. Although they are the most accurate determination of these values, they are still affected by various uncertainties, depending on the different assumptions made to perform the photo-ionization models. These uncertainties can potentially be very large and are not fully quantified by the error bars shown in Figure~\ref{fig:comparisons}. As an example, the  COS-Halos collaboration \citep{werk13,werk14,tumlinson13} analysed with similar photo-ionization models and slightly different assumptions \citep[see][]{prochaska17} 5 of the galaxies of our sample, finding column densities that can be 1 or 2 orders of magnitudes different from the COS-LRG estimates. Because of these uncertainties in the data, we have not included the column densities in the MCMC fit.
We can note however that our predictions are not inconsistent with the observations. Interestingly, with our models we can also make predictions about the density profile in the outer part of the halos: we will discuss the implications in Section~\ref{compardisc}.\\
We conclude that, with the correct choice of parameters, models of clouds that are starting from the virial radii of the halos, infalling at a cosmologically motivated rate and then interacting with the hot coronal gas, can successfully reproduce the observed kinematics, number of absorbers and column densities of the cool CGM around early-type galaxies. This is our first and main result.\\
In Section~\ref{motiondescription} we analyze the motion and the properties of our model clouds and we discuss the physical scenario that these particular models describe. 
\subsection{Description of the cloud motion}\label{motiondescription}
\begin{figure}[ht!]
   \includegraphics[width=1\linewidth]{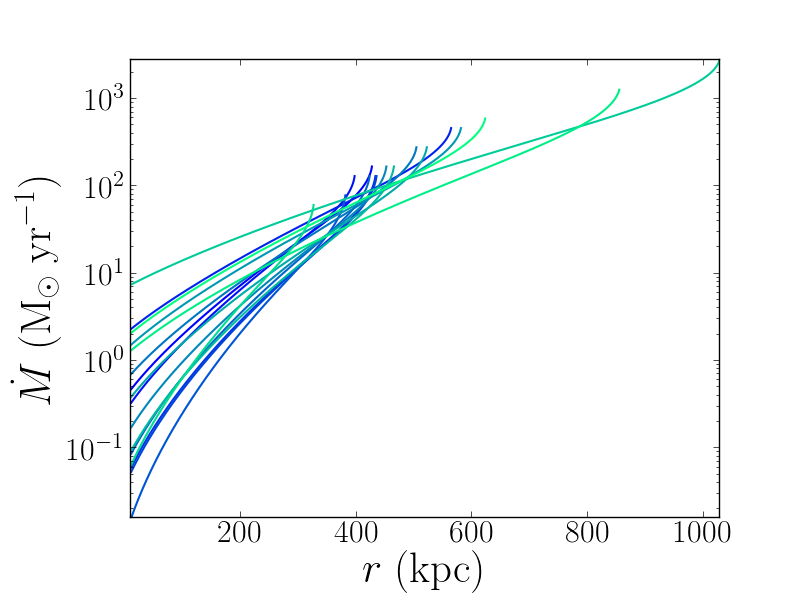}
   \caption{Same as Figure~\ref{fig:resprof}, but for the total mass flux profile of the cool CGM.}
              \label{fig:resprof1}%
    \end{figure}

Using the median parameters of Table~\ref{tab:mcmcres} we can have an indication on the properties of the best models that reproduce the observations and we can analyze in detail the motion of the infalling clouds.  
In Figure~\ref{fig:resprof} and \ref{fig:resprof1} we report our results on the cool CGM properties, all with respect to the intrinsic distance $r$ from the central galaxies.
For the moment we do not discuss the possible origin of the cool clouds (Section~\ref{origin}), but we just describe their behavior throughout their infall.
The different blue-tone lines in Figure~\ref{fig:resprof} and \ref{fig:resprof1} represent the results obtained for the 16 different galaxies that we are modeling. Since we are describing all the galaxies with the same model, the general trends of the cool CGM properties are similar in all the objects and the arising physical picture is the same for all the galaxies of our sample, with small variations due to the different virial masses and radii.\\
In the top left panel of Figure~\ref{fig:resprof} we show the cloud velocity profiles. The clouds start at the virial radii with very low velocities, they are accelerated by the gravitational force and then, as already explained in Section~\ref{model}, they are decelerated by the drag force and the other hydrodynamical interactions. We stress that it is this hydrodynamical deceleration that allows the resultant line-of-sight velocity distribution to be consistent with the observed one (see left panel of Figure~\ref{fig:comparisons}).\\
With our analysis we can also estimate, only using our analytical description of these systems, the mass and size of the cool CGM clouds, which were so far largely uncertain \citep{stocke13,werk14}. The top right and bottom left panels of Figure~\ref{fig:resprof} show the evolution of these two quantities and the direct effects of the evaporation and the pressure equilibrium on the clouds. The profiles in the top panel show that the clouds start at the virial radii of the galaxies with a mass of  $\sim10^{5}\ \rm{M}_{\odot}$, then they progressively lose mass due to the interaction with the corona. In the central region they have lost more than 99$\%$ of their mass and have masses of about $10^{2}\ \rm{M}_{\odot}$. Also the radii of the clouds (bottom panel), obtained from equation~\eqref{eq:radii}, strongly decrease with time, starting from $r_{\rm{cl}}\sim 1$ kpc at the virial radius and reaching sizes of a few pc in the central regions, due to the effects of both the cloud evaporation and the pressure equilibrium (in the central regions the corona and cloud densities are higher). The conclusion is that in the central regions the cool CGM clouds are much smaller and also less massive, because most of their mass is evaporated in the hot corona.\\
From the bottom right panel of Figure~\ref{fig:resprof}, which shows the cumulative mass profiles of the cool gas, we can infer how the mass loss of the clouds influences the distribution of the cool CGM mass throughout the halo. Most of the cool gas mass is concentrated in the external regions of the halos, because of the volume effect (their volume is much larger than the one of the internal parts), but also because the clouds are more massive. In fact, because of the cloud evaporation, the decline of the mass profiles is much steeper than what expected without evaporation and the inner regions are almost devoid of cool gas. While on average the total mass of the cool medium within the galaxy halo is $2.5\times10^{11}\ \rm{M_{\odot}}$, the cool gas mass within $r=160$ kpc (highlighted by the vertical dashed line) is only $\sim1\%$ of the total. Therefore, we are describing with our models a system where most of the cool CGM gas is not reaching the central regions and is not accreting onto the galaxy. This is also shown by the cool gas accretion rate, reported in Figure~\ref{fig:resprof1}: starting from a cosmological rate at the virial radius, the accretion in the inner parts is reduced to less than $1\%$ of the initial value.
   \\ 
With this result we are therefore in agreement with the observations of a great amount of cool CGM in these halos and we also go in the direction of explaining why this large reservoir of cool gas is not feeding the star formation of these thus passive central early-type galaxies. We discuss this further in~\ref{cloudsurv}.\\
\section{Discussion}\label{discussion}
With our analysis we have found that infall models of cool clouds can successfully describe the COS-LRG observations of the cool CGM around massive ETGs. We presented our findings in 
Section~\ref{resdisc}. In this Section we discuss all the possible limitations of our analytical models (Sections~\ref{cloudsurv} and~\ref{explainaccr}), the comparison with other observations (Section~\ref{compardisc}) and the implications of our findings for the cool gas state, origin and fate (Sections~\ref{presseq} and \ref{origin}).
\subsection{Cloud survival}\label{cloudsurv}
We have seen in Section~\ref{motiondescription} that with our models we describe a scenario where, due to the evaporation of the CGM clouds, only a very small part of the cool gas accreting into the galaxy halos is reaching the inner regions. Nevertheless, we also find that this small amount of cool gas (on average a few $\rm{M}_{\odot}\ \rm{yr}^{-1}$) is reaching the central ETGs, possibly feeding the galaxy star formation at a similar rate.
In this Section we discuss whether this problem may be alleviated by additional effects that we have not taken into account, focusing in particular on the survival time of the cool clouds.\\
As already explained, the analytical model of Section~\ref{model} is only an approximation of all the hydrodynamical interactions taking place in the CGM. As a consequence, the clouds described in our treatment can lose mass, but they cannot completely evaporate in the hot gas or be destroyed by the cloud-corona interactions, which is instead what we expect in reality and what we observe in high-resolution simulations \citep[e.g.][]{heitsch09}. Therefore, it is important to estimate the time after which the clouds are probably completely evaporated in the surrounding hot medium. We consider as an estimator of the cloud survival time the conduction time, which is the time needed for the thermal conduction to be efficient and therefore for the cloud to be entirely evaporated in the corona. In our massive galaxy halos, where there is a high temperature gradient between the different gas layers (the coronal gas is almost three orders of magnitudes hotter than the cool clouds) we expect the thermal conduction to be particularly efficient and to play a major role in the evaporation of the clouds \citep{armillotta17}.\\ The conduction time is defined as \citep{spitzer62}
\begin{ceqn}
\begin{equation}\label{eq:condtime}
t_{\rm{cond}}=\frac{n_{\rm{cool}}k_{\rm{B}}r^2_{\rm{cl}}}{fk_{\rm{sp}}}
\end{equation}
\end{ceqn}
where $f$ is the suppression factor due to the magnetic field (e.g. \citealt{chandran98}) and:
\begin{ceqn}
\begin{equation}\label{eq:spitzer}
k_{\rm{sp}}=\frac{1.84\times 10^{-5}T^{5/2}}{\ln \Psi}\ \rm{erg}\ \rm{s}^{-1}\ K^{-1}\ \rm{cm}^{-1},
\end{equation}
\end{ceqn}
is the heat conduction coefficient, where $\ln \Psi$ is the Coulomb logarithm ($\approx30$) and $T$ is the temperature in K of the hot gas. We estimated the mean value of this time for a cloud at the virial radius with $m_{\rm{cl,start}}\sim10^5\ \rm{M}_{\odot}$ (see Table~\ref{tab:mcmcres}) and $f=0.1$, obtaining $t_{\rm{cond}}\sim200\ \rm{Myr}$.
This has to be compared to the mean infall time of a cloud in our best models, calculated using the formula
\begin{ceqn}
\begin{equation}\label{eq:inftime}
t_{\rm{fall}}=\int^0_{r_{\rm{vir}}}\frac{dr}{v(r)}\ .
\end{equation}
\end{ceqn}
Averaging this time over the seven galaxies of our sample, we obtained that $t_{\rm{fall}}\sim3.7\ \rm{Gyr}$, much longer than the estimated conduction time. This means that the clouds cannot survive their entire journey from the virial radius to the central galaxy and they most likely completely evaporate in the hot gas at relatively large distances from the center of the halos. Therefore, the cool CGM is not feeding the central galaxies, thus explaining the quiescence of these objects. However, the comparison between these two times is not straightforward, since there are uncertainties on both quantities. The derived value of the conduction time is in fact dependent on the magnetic field, whose intensity and orientation is unknown. If the magnetic field suppression is larger than what we have considered in our estimate, the suppression factor $f$ in equation~\eqref{eq:condtime} would be lower and therefore the conduction time would be longer than what we have found. On the other hand, since the clouds become smaller with time, the conduction is more efficient in the inner regions of the halos and therefore our estimate is an upper limit on the survival time of the clouds. Finally, there are also uncertainties on the infall time obtained through equation~\eqref{eq:inftime}, since we are considering in our analysis only a radial infall. Adding any tangential component would probably increase the final estimate of the total cloud infall time. 
Based on these considerations, we consider very likely that the clouds cannot survive for the entire infall time, although it is hard to define precisely the region of the halo where they completely evaporate.\\
The kinematic distribution of our model is arising from clouds distributed all over the halos, while we have argued with the last estimation that the central regions are probably devoid of cool gas. To prove that this property does not affect our final findings, we have performed again the MCMC analysis, with the same likelihood expressed in Section~\ref{comparison} and the same parameter space outlined in Section~\ref{mcmcres}, but with the central regions of the halos forced to be completely devoid of clouds. To do this, we have defined the destruction radius $r_{\rm{destr}}$ as the inferior limit of the integral in equation~\eqref{eq:totnumber} in place of 0 and as the smallest intrinsic radius where the clouds are present in our distribution. In this way we can assess how much the cloud destruction influences the final results. We have run a second MCMC, with $r_{\rm{destr}}=r_{\rm{vir}}/2$, obtaining the results reported in Table~\ref{tab:mcmcnew}. We can see how the best-fit parameters are consistent within the errors with the ones found with the first MCMC, reported in Table~\ref{tab:mcmcres}. The physical scenario is the same described in Section~\ref{motiondescription}, but with the cool clouds by construction not reaching the central galaxies. Also with this kind of model we are able to successfully reproduce the 2 observational constraints. In Figure~\ref{fig:CDF} we show the comparison between the cumulative distribution functions of the velocity distributions of the models and the observations: the kinematic distribution of the last model is perfectly consistent both with the result of our main model (Section~\ref{resdisc}) and with the observations, with a number of clouds equal to $44\pm 11$. The reason why the results of our model did not change with this last modification is that, even without imposing the cloud destruction, the vast majority the clouds observed by our synthetic observation analysis (see Section~\ref{comparison}) are residing in the halo external regions. The radii of the external absorbers are in fact up to 2 orders of magnitudes larger than the internal ones (bottom left panel of Figure~\ref{fig:resprof}), implying much higher cross sections and therefore higher chances to be observed.\\
We can finally conclude that our model results, although we cannot directly simulate the complete evaporation of the cool clouds, provide a coherent picture where the pressure-confined cool CGM absorbers are: 1) infalling toward the center, 2) slowed down by the coronal drag force and 3) completely destroyed by the interactions with the hot gas before reaching the galaxy, thus not feeding any central star formation.\\
A similar scenario was discussed by \cite{huang16} to explain the kinematics of cool CGM clouds around massive elliptical galaxies: they argue that the observed low velocities could be reproduced by clouds with $m_{\rm{cl}}\sim5\times10^4\ \rm{M_{\odot}}$, slowed down by the effect of the hot gas drag force, though they do not assess this possibility with detailed dynamical models, as we have developed with our work. Also \cite{chen18} discuss the same physical scenario for the cool CGM, arguing that most of the clouds with $m_{\rm{cl}}<10^6\ \rm{M_{\odot}}$ cannot reach the center if they originate at large distances from the galaxies, in agreement with our final findings. Our work is the first attempt to develop a coherent dynamical model to describe the CGM around massive ETGs. Similar semi-analytical models for the cool circumgalactic medium have recently been developed by \cite{lan18}, though they focus on $10^{12}\ \rm{M}_{\odot}$ halos around star-forming galaxies, for which they find that an outflow model can qualitatively reproduce several CGM observed properties. Although our approaches are similar, we have a more realistic description of the clouds kinematics, since we are including the drag force in our setup, which has a great impact on the final results and is instead neglected in \cite{lan18}, and we directly compare our results with the full observed velocity distribution of the cool gas clouds.
\subsection{Pressure equilibrium}\label{presseq}
A fundamental assumption that we made for the cool CGM physical state is the pressure equilibrium between the cool absorbers and the coronal gas. This is physically justified, since we expect any pressure imbalance with the ambient medium to be erased on a cloud sound crossing time-scale, which is roughly $\sim 10$ Myr and therefore much smaller than the dynamical time.\\ It is hard to compare our intrinsic densities with the COS-LRG data, because of the already mentioned uncertainties in the observations, due to photo-ionization modelling and, most importantly, to projection effects. However, \cite{zahedy19} find, through their photo-ionization analysis, that the cool gas is likely in pressure equilibrium with the hot halo, in agreement with our assumption of clouds pressure-confined by the corona. We conclude that, in addition to being physically motivated, this assumption is also supported by observations (although earlier studies had been suggesting different results, see \citealt{werk14}).
We plan to investigate these aspects further, including explicit photo-ionization modelling, in future works.
{ 
\renewcommand{\arraystretch}{1.5}
 \begin{table}[htbp]
\begin{center}
 \caption{MCMC results of the two additional models.}\label{tab:mcmcnew}
\begin{tabular}{*{4}{c}}
\hline  
\hline
 Model & $\log\ (m_{\rm{cl,start}}/\rm{M}_{\odot})$  & $\alpha$ & $f_{\rm{accr}}$\\
      &      &   $(\rm{Gyr}^{-1})$         &\\
\hline 
$r_{\rm{destr}}=r_{\rm{vir}}/2$ &$4.88\raisebox{.2ex}{$\substack{+0.23 \\ -0.22}$}$ & $1.71\raisebox{.2ex}{$\substack{+0.31 \\ -0.33}$}$ & $1.29\raisebox{.2ex}{$\substack{+0.20 \\ -0.21}$}$ \\
$r_{\rm{start}}=r_{\rm{vir}}/2$& $4.91\raisebox{.2ex}{$\substack{+0.25 \\ -0.26}$}$ & $2.40\raisebox{.2ex}{$\substack{+0.52 \\ -0.55}$}$ & $1.20\raisebox{.2ex}{$\substack{+0.19 \\ -0.18}$}$ \\
\hline
\end{tabular}
\end{center}
\tablefoot{50th percentiles (with errors given by the 32nd and the 68th percentiles) of the posterior distributions of the three parameters obtained with the MCMC fits performed for models with $r_{\rm{destr}}=r_{\rm{vir}}/2$ (Section~\ref{cloudsurv}) and $r_{\rm{start}}=r_{\rm{vir}}/2$ (Section~\ref{compardisc}).}
   \end{table}
   }
\subsection{Comparison with other observations}\label{compardisc}
The results of this work are obtained from the comparison of our model predictions with the observations of the COS-LRG collaboration, which analyzed both metal and hydrogen CGM absorption lines with a very high spectral resolution, allowing us to have detailed data on the kinematics of the cool clouds around 16 massive ETGs (Section~\ref{obs}). A comparable analysis has been done by the COS-Halos program \citep{werk13,werk14,tumlinson13,prochaska17}, a larger survey of the cool CGM around galaxies with a broad stellar mass range. Five of the galaxies analysed by this program are also part of the COS-LRG sample. We have applied our models also on this subsample of 5 objects, using only the kinematic data of the COS-Halos collaboration. The results of the MCMC analysis are totally consistent with the ones showed in this paper.\\
However, these data sets have also some limitations, mostly due to the lack of observations at projected radii larger than 160 kpc (in both samples the projected distance of the galaxy from the background quasar is always smaller than 160 kpc). It is important therefore to compare our findings with observations at larger projected distances. \cite{zhu14} and \cite{huang16} have observed MgII absorptions from cool CGM around large samples of ETGs out to very large distances from the central galaxies and they both found that the cool gas has a kinematics consistent with what is found by COS-LRG and COS-Halos. Their data have not been included in our analysis, because the resolution of these observations is not sufficient to disentangle the different clouds along a single line of sight. Also, they observe only the MgII, which could be not representative of the behavior of the whole gas. However, it is interesting to compare our predictions with their results.\\
As seen in Section~\ref{mcmcres}, with our model we can make predictions on the cool CGM behavior at large distances, in particular for the projected profile of the total hydrogen column density. The result of our best model is reported as a blue curve in the right panel of Figure~\ref{fig:comparisons}: the column density is only slightly declining at large distances from the galaxies. This behavior is expected because our model predicts that most of the cool gas is concentrated in the halo external regions, es explained in Section~\ref{motiondescription}. Therefore, the amount of gas observed by a line of sight at a large impact parameter is almost equal to the one observed at low projected distances from the galaxy and the resulting projected profile is nearly flat. By construction, the column density of each galaxy goes to zero at the virial radius, beyond which, in our modeling, there is no cool gas. The gradual decline of the column density profile is therefore also due to the fact that different galaxies live in different dark matter halos, with the small bumps corresponding to the different virial radii (reported in Table~\ref{tab:galprop}), where the column density of each galaxy goes to zero.\\
\cite{huang16} show that there are strong MgII absorptions up to 500 kpc from the central galaxies, with the covering fraction declining at projected radii larger than 100 kpc. \cite{zhu14}, who analyze stacked spectra, observe MgII absorptions till and beyond the galaxy virial radii, finding that the total MgII surface density is decreasing with the projected distance. Both works are therefore consistent with our findings of cool absorbers still present at very large distances from the central galaxies and the decrease in the absorption strength, which is less than 2 order of magnitudes from the center to the virial radius, is roughly compatible with the decline of our total hydrogen column density.\\
Recently \cite{berg18} have analyzed the HI in the CGM around 21 massive ETGs, with sightlines within $\sim 500$ kpc from the central object. They find a covering fraction of the cool gas that is decreasing going from the central regions to high impact parameters. Our models also predict a decline in the projected profile of the covering fraction that is consistent with these observations, although it is hard to make a quantitative comparison between observations and model predictions. Since \cite{berg18} have not provided kinematic estimations, we have not included these data in our analysis.
   \begin{figure}[ht!]
   \includegraphics[width=1\linewidth]{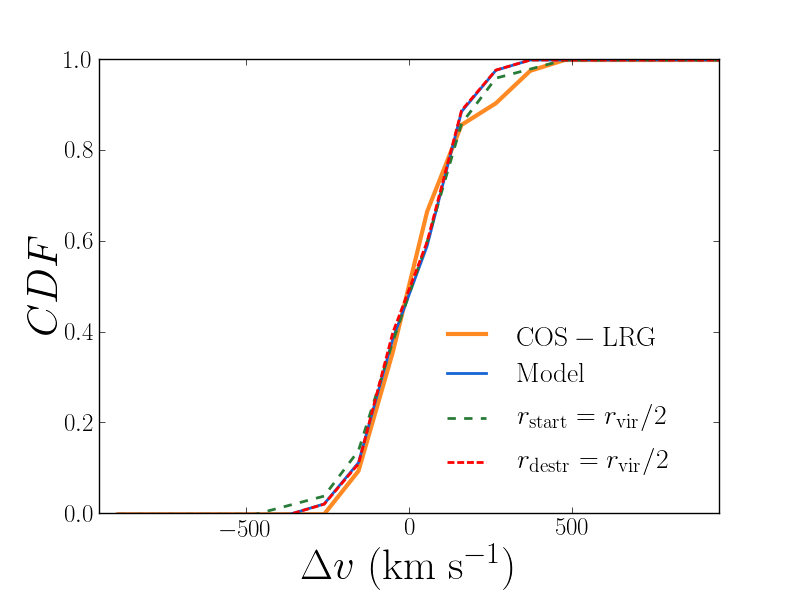}
   \caption{Cumulative Distribution Functions of the line-of-sight velocity distributions of the observations (orange curve) and of our models. In blue we report the result of our main model (Section~\ref{resdisc}), while the dashed curves represent models with $r_{\rm{start}}=r_{\rm{vir}}/2$ (green, dashed) and $r_{\rm{destr}}=r_{\rm{vir}}/2$ (red, small dashed).}
              \label{fig:CDF}%
    \end{figure}\\
There is also the possibility that the cool clouds start from distances smaller than the virial radius (see Section~\ref{altor}) and in this scenario the steepness of the column density and the covering fraction profiles would be different, with very low values at large distances. With our current dataset we cannot directly constraint the cloud starting radius, since we are using observations till 160 kpc. Therefore, instead of considering the starting radius as an additional free parameter of our models, we have run an MCMC analysis as in Section~\ref{comparison}, but fixing the starting radius at the value of $r_{\rm{start}}=r_{\rm{vir}}/2$. We present the results of this fit in Table~\ref{tab:mcmcnew}: the best parameters are consistent with our previous findings, with a physical scenario similar to the one described in Section~\ref{motiondescription}. Also, the line-of-sight velocity distribution and the number of clouds ($49\pm 13$) resulting from this model are consistent with the observations, as shown by the CDFs comparison in Figure~\ref{fig:CDF}. Therefore, even models with smaller starting radii can successfully describe the cool CGM properties. 
Future high-resolution observations of cool CGM absorbers at impact parameters comparable to the virial radius will help distinguishing between different scenarios.
\subsection{Uncertainties in the accretion rate}\label{explainaccr}
In Section~\ref{mcmcres} we have shown that the best models in reproducing the observations have a mass accretion rate which is slightly (1.2 times) higher than the prediction of cosmological models, while in principle we would not expect the cool mass accretion to be greater than the cosmological rate. The reasons for this unexpected value are probably due to the simplifications of our models. The first reason could be the simplified structure of our model clouds. They are in fact spherical, with their radius becoming smaller as they fall throughout the halos. Although this is probably a good approximation for the head of the clouds, hydrodynamical simulations show that the interactions with the hot gas create a tail of partially mixed gas which can be much more extended than the main head. We cannot model this effect with analytical descriptions, but we argue that it would mainly have an effect on the number of clouds observed by the synthetic observations of our models. The cross section of the clouds will indeed be larger and we would synthetically observe more clouds than what we are doing with our current models. Therefore, to observe the same number of clouds we would need a lower total accretion of cool gas, in agreement with the prediction of the cosmological models. Note that this effect would likely have also an impact on the shape of the model velocity distribution. With our simplification of the structure of the clouds, we are indeed underestimating mostly the cross section of the slowest clouds, that had more interactions with the corona. Therefore we expect, from a more thorough analysis, to observe more clouds with very low velocities, populating the very central region of the model velocity distribution (see the left panel of Figure~\ref{fig:comparisons}) and alleviating the tension in that bin between the observations and our predictions.\\
Another reason why we are probably overestimating the accretion rate is instead given by our assumption of isotropic inflow onto the galaxy halos. As explained in Sections~\ref{accrrate} and \ref{comparison}, to calculate the mass accretion rate of the cool gas we assumed that the cool gas is inflowing isotropically, starting from the virial radius. However, the accretion could be anisotropic, with the cool medium inflowing through filaments toward the galaxies (cool gas filamentary accretion is common in high-resolution cosmological simulations, see \citealt{nelson16}). In this scenario, the filaments would not fill the entire volume of the halos and the total accretion of gas would be different than what predicted by our model.
However, the low number statistics does not allow us to constraint the real geometry of the gas inflow and to understand whether and how much we are overestimating the cool CGM accretion with our simplified, isotropic models. We leave a more detailed analysis about the geometry of the gas accretion for future works.
\subsection{Origin and fate of the cool absorbers}\label{origin}
In the previous Sections we have focused on the physical description of the cool absorbers behavior throughout the halos. In this Section we instead analyze the possible origin scenarios for the cool CGM, as well as their consequences on the galaxy evolution. 
The aim of this Section is only to speculate on the different possible pictures that can arise from our models, leaving a detailed analysis of the cool CGM origin and fate for further works.
\subsubsection{External origin}\label{cosmfil}
Our favourite interpretation for the origin of the cool CGM is that it is coming from the accretion of external gas. As explained in Section~\ref{model} in fact, cosmological models predict that galaxy halos acquire gas at a rate described by equation~\eqref{eq:cosmaccr} and our assumption is that the cool CGM is originated by this accretion of external gas: in our models we set indeed the cool gas accretion rate in order to be consistent with the cosmological one. In this picture, the filaments of low-metallicity intergalactic gas accrete into the galaxy halo and form the cool clouds that fall down through the halos as described by our modeling.\\
This assumption is motivated by the COS-LRG analysis of the cool CGM metallicities \citep{zahedy19}: they find in fact that about half of the galaxies in their sample host metal poor clouds, consistently with our picture of accretion of external gas, which is expected to have low metallicities \citep[see][]{lehner16}. Also a recent work by \cite{chen19} states that very low metallicity gas is present in the halos of massive ETGs, probably coming from the accretion of intergalactic medium. The scenario of external origin has more difficulties to explain the observations of absorbers with higher metallicities, that are common in the COS-LRG survey and also in previous surveys like COS-Halos. \cite{prochaska17} find indeed that the CGM absorbers in the whole COS-Halos sample span a broad range in metallicities, with a median value around 30\% solar. A way to produce these metal-rich absorbers is to invoke internal origins for the clouds, which we will discuss in Section~\ref{altor}.\\
Another way to explain the observation of a broad range of metallicities is with the destruction of the cool accreting filaments.
It is possible in fact that the clouds are formed by the fragmentation of the filaments and their mixing with the corona, explaining higher metallicities than what expected in the IGM. Also, in this scenario, the filaments are slowed down by these processes and it is reasonable to believe that the resulting clouds will start with low velocities, as indicated by our results\footnote{The fragmentation of the filaments could happen also at radii larger than the virial radius, for example at the virial shock radius ($\approx1.25$ $r_{\rm{vir}}$, see \citealt{nelson16}). In this scenario, the filaments are fragmented outside the galaxy halo and they enter the virial radius already in the form of slow-moving clouds.} (see Appendix~\ref{invel}).
Due to these considerations, we consider the external accretion consistent with our results and the most likely picture for the cool CGM absorbers around ETGs.
\subsubsection{Alternative origins}\label{altor}
A plausible alternative origin for the cool curcumgalactic gas in the halos of massive ETGs would be from the stripping of ISM of satellite galaxies by tidal or ram pressure forces. Signatures of cool gas coming from this type of interactions have been observed directly in emission in different massive halos \citep[e.g.][]{epinat18,johnson18}. The cool gas stripped by the satellite could eventually fall down towards the central massive galaxy, in a journey compatible with the motion described by our model. Although the accretion from external gas remains our favourite scenario, we do not reject this last picture as a possibility for the cool CGM origin.\\
Another scenario that can give rise to cool clouds falling through the halos is related to the presence of outflows from the central galaxy, which we have not considered so far in our treatment. This neglect is justified by the nature of the galaxies that we are studying, that are by definition in a quiescent state and should therefore not experience strong outflows of gas. However, although they are currently quiescent, there could have been outflows in the past of these objects, due to either star formation bursts or AGN activity. AGN are indeed known to drive massive multiphase outflows \citep[e.g.][]{greene12} and cool, high velocity gas is often observed in the CGM of quasars \citep[e.g.][]{bowen06,prochaska13,johnson15}. It is possible to think that, as in the recycling scenario (see Section~\ref{intro}), the outflows due to recent star formation or AGN episodes have reached the external parts of the halos, cooled and eventually fallen down again toward the center. In this picture we are observing the last part of this cycle, composed by the infalling cool CGM clouds, which in this case have an internal origin.
This would mean that we are not observing a continuous accretion but a sporadic event of infall due to a recent outflow episode happened in the central galaxies. Although this picture would explain the high metallicities observed in part of the cool CGM \citep{prochaska17}, all our galaxies should have experienced an outflow burst at similar times, with comparable timescales and we consider this scenario unlikely. Thus, the recycling is most probably not the main driver of the cool CGM formation around early-type galaxies.\\
The last possibility is that the clouds are condensing out of the hot gas due to thermal instabilities, developing a multiphase halo. Multiphase gas has indeed been observed in both nearby clusters and elliptical galaxies \citep{voit15a,voit15b} and the same origin scenario has been proposed also for the cool CGM around massive ETGs by different observational work \citep[see][]{huang16, zahedy19}. This picture is motivated by various theoretical works \citep[e.g.][]{sharma12,mccourt12,voit18}, which predict that the instabilities can develop at a cooling radius $r_{\rm{c}}<r_{\rm{vir}}$, where the cooling time is comparable to the dynamical time of the gas.
In this scenario, clouds are created within the galaxy halos with very low velocities, (which may be of the order of the turbulence of the hot coronal gas, see \citealt{voit18}) and then start to fall down toward the central galaxies, consistently with our main findings. However, it is so far not clear if such instabilities can spontaneously develop in the hot corona \citep[e.g.][]{binney09,nipoti14} and therefore it is difficult to establish whether this scenario could be an important contributor to the cool CGM.\\
We conclude then that the bulk of the cool circumgalactic gas is likely coming from the accretion of the external IGM, although the different internal processes outlined above may originate part of the observed cool clouds, also explaining the broad range in the observed metallicities.
\subsubsection{Fate of the cool gas}
Many works in the last decades have studied the fate of the accretion of cool gas from the IGM onto galaxy halos, both with analytical prescriptions and cosmological simulations \citep[e.g.][]{birnboim03,ocvirk08,dekel09,nelson13,nelson16}. The general picture is that, in low mass DM halos, filaments can penetrate through the halo reaching the central galaxy, while in the most massive objects the filaments are shock heated to the galaxy virial temperature and are unable to feed the central galaxy. In this scenario, the CGM clouds that we are studying are the products of the fragmentation of the cool gas filaments. In agreement with the general picture outlined above, we predict that these cool gas components are not able to survive their journey and reach the central galaxy (see Section~\ref{cloudsurv}), but they instead evaporate in the hot corona at large distances from the centre. We therefore believe that these absorbers cannot be straightforwardly linked with the observations in less massive halos of High Velocity Clouds (HVCs, see \citealt{putman12}), clouds of cold gas observed mostly in HI both around the Milky Way and nearby star-forming galaxies. HVCs reside in fact primarily within 10 kpc from the central galaxy disk and their origin is still debated \citep{fraternali15}. On the contrary, we expect our CGM clouds around ETGs to be completely destroyed by the interactions with the hot gas at much larger distances. The fate of the cool CGM, by the evaporation of the clouds, is therefore to increase the mass of the hot corona.\\ 
From Figure~\ref{fig:resprof1} we can infer the rate of cool gas accreting onto the galaxy halos: on average, the gas accretion at the virial radius is about 500 $\rm{M}_{\odot}\ \rm{yr}^{-1}$. If it is coming from the cosmological inflow, this amount of cool gas is continuously injecting baryons in the galaxy halo, by evaporating into the hot coronal gas. 
At this rate, (which could be slightly lower, due to the considerations of Section~\ref{explainaccr}) the mass of the corona would be doubled in about 2 Gyr. 
This may be related to the current and future location of the so-called missing baryons \citep{bregman07,mcgaugh08}, although a more thorough discussion of these aspects is outside the scope of this work.
\section{Summary and conclusions}\label{concl}
In this work we carried out a detailed dynamical analysis of the cool circumgalactic medium around massive early-type galaxies at low-redshift, in order to explain the presence of a substantial amount of cool gas around quiescent galaxies and the narrowness of the observed velocity distribution of the cool CGM clouds.\\
In particular we focused on the observations of the COS-LRG collaboration around 16 massive ETGs. We developed semi-analytical models of cool clouds infalling from the external regions of the halos to the central galaxies, approximating the hydrodynamical interactions of the clouds with the hot corona with analytical parametrizations. These models aim to reproduce the two observational constraints given by the kinematic distribution and the number of the cool clouds. Our model has three free parameters that account for the properties of the infalling clouds and for their interaction with the ambient medium. We constrained these free parameters on the available data with an MCMC analysis.\\
From our results we draw the following conclusions:
\begin{enumerate}
\item models of cool CGM clouds infalling through the galaxy halos at a cosmologically motivated rate can successfully reproduce the observations, explaining the number of observed clouds and their line-of-sight velocity distribution. These results are also consistent with the total hydrogen cool CGM column densities reported by COS-LRG;\\
\item our best models describe clouds with an initial mass $m_{\rm{cl}}\sim10^5\ \rm{M}_{\odot}$ at the virial radius, that during their infall lose more than $99\%$ of their mass due to the hydrodynamical interactions with the hot coronal gas, implying that the internal regions of the halos are almost devoid of cool gas;\\
\item despite the uncertainties given by our approximation in the hydrodynamical treatment, we can conclude that the cool CGM clouds most probably cannot survive the journey but will completely evaporate into the hot corona as the timescale of thermal conduction is much shorter than the dynamical time. Thus the cool gas will probably not accrete onto the central galaxies, providing an explanation of their quiescence.
\end{enumerate}
The results of this paper represent a step forward in solving the puzzle of the existence of large unused reservoirs of cool circumgalactic gas around quiescent galaxies. Future improvement of the observational data and ad-hoc hydrodynamical simulations could reduce the uncertainties in our current description. We also plan to apply similar models to larger samples of data and in particular to star forming galaxies, in order to study the impact of the CGM kinematics and dynamics on the evolution of different kinds of galaxies.
\begin{acknowledgements}
The authors would like to thank the referee Sean Johnson for a very constructive report and the helpful comments and suggestions. We are also very grateful to Lorenzo Posti and Antonino Marasco for the useful discussions and suggestions that significantly helped to improve this work. GP acknowledges support by the Swiss National Science Foundation, grant $\rm{PP00P2}\_163824$.
\end{acknowledgements}
\bibliographystyle{aa} 
\bibliography{biblio}
\begin{appendix}
\section{Geometry of the system}\label{geometry}
In Section~\ref{comparison} we have described how we compared our results with the observations, using synthetic observations of random populations of clouds. Here we report in detail the geometry used to infer the number of 'observed' clouds and their line-of-sight velocities. We created for each galaxy a 3-dimensional distribution of clouds over a sphere with radius $r_{\rm{vir}}$ (see Table~\ref{tab:galprop}), representing the virial halo. As already explained in Section~\ref{comparison}, the total number of clouds for each halo is calculated using equation~\eqref{eq:totnumber}, while the intrinsic distance of each cloud from the galaxy is found using the probability density function expressed by~\eqref{eq:pdf}. We then associated to each cloud two other coordinates $\theta$ and $\phi$. These last two coordinates are chosen to have the clouds uniformly distributed over the sphere: we created a random distribution of the angle $\phi$ uniformly distributed between 0 and $2\pi$ and a random distribution of $\cos \theta$ uniformly distributed between -1 and 1, with $0<\theta<\pi$. Each cloud has therefore a specific position in the halo, expressed by the coordinates (r, $\theta$, $\phi$).
   \begin{figure}[h!]
   \includegraphics[width=1\linewidth]{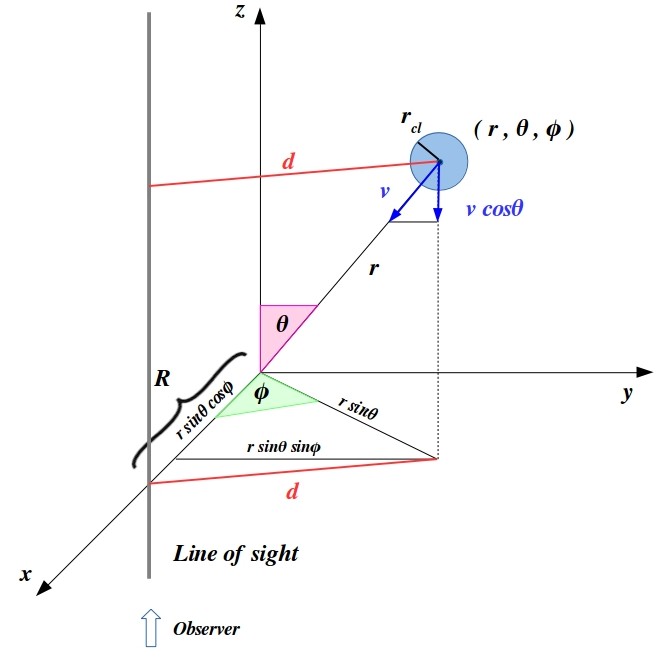}
   \caption{3-dimensional representation of the synthetic observation of a generic cloud, as performed in our model.}
              \label{fig:geom}%
    \end{figure}\\
           \begin{figure*}[hb!]
   \centering
   \includegraphics[clip, trim={0 0 0cm 0}, width=14.9cm]{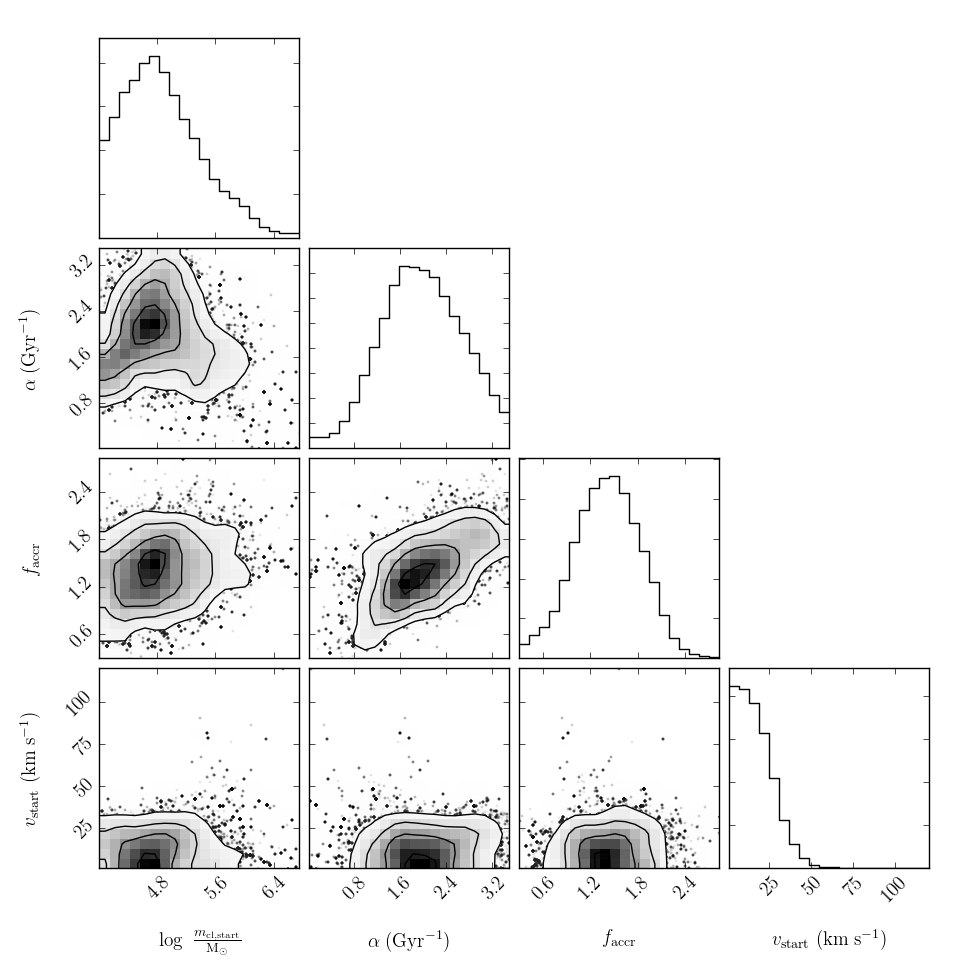}
   \caption{Corner plot with the results of the MCMC analysis described in Appendix~\ref{invel}, where also the starting velocity of the clouds is added as a fourth free parameter. Both the one and two dimensional projections of the posterior probabilities of the four free parameters are shown.}
              \label{fig:MCMC_4par}%
    \end{figure*}
Figure~\ref{fig:geom} shows the method used to perform the synthetic observations. Here, the line of sight is given by the line at $x=R$ in the plane $\phi=0$, where $R$ is the distance of the galaxy from the sightline, reported for each object in Table~\ref{tab:galprop}. The cloud in (r, $\theta$, $\phi$) is observed if the distance $d$ of this point from the line of sight is smaller than the cloud radius, expressed by
\begin{ceqn}
\begin{equation}\label{eq:clouddist}
d<r_{\rm{cl}}\ ,
\end{equation}
\end{ceqn}
where 
\begin{ceqn}
\begin{equation}\label{eq:clouddist1}
d=\sqrt{\left( R-r_{\rm{i}}\sin{\theta}\cos{\phi} \right)^2+(r_{\rm{i}}\sin{\theta}\sin{\phi})^2}\ .
\end{equation}
\end{ceqn}\\
If the cloud is observed, i.e. if condition~\eqref{eq:clouddist} is true, we calculated its line-of-sight velocity, as shown in Figure~\ref{fig:geom}, through
\begin{ceqn}
\begin{equation}\label{eq:vlos}
v_{\rm{los}}=v\cos{\theta}\ .
\end{equation}
\end{ceqn}
and we added it to the model velocity distribution to be compared with the COS-LRG observations.
\section{Cloud initial velocity}\label{invel}
In Section~\ref{model} we have described the construction of our semi-analytical models and we have argued that the starting velocity of the clouds ($v_{\rm{start}}$, the velocity of the clouds at the virial radius of the galaxy) is needed to solve equation~\eqref{eq:motion}. In this Appendix we assess models where $v_{\rm{start}}$ is not fixed, but free to vary between different values. In order to do this, we have run an MCMC analysis as in Section~\ref{comparison}, but adding $v_{\rm{start}}$ as a fourth free parameter, with a flat prior from 0 to 350 km $\rm{s}^{-1}$. Figure~\ref{fig:MCMC_4par} shows the results of this analysis, the one and two dimensional projections of the posterior probabilities of the four parameters. The distributions of the first three parameters are in strong agreement with the ones in Figure~\ref{fig:MCMC}, meaning that adding the initial velocity as a free parameter does not change the final results.\\
The four bottom panels of Figure~\ref{fig:MCMC_4par} represent the posterior distributions regarding $v_{\rm{start}}$. We can note that models with clouds starting from the virial radius with high velocities are strongly disfavored by the MCMC analysis and that the distributions are instead peaked on very low values. This means that the best models in reproducing the COS-LRG observations have clouds with initial velocities close to zero. This condition is indeed needed by the models to reproduce the central region of the line-of-sight velocity distribution, which shows an excess of clouds with very low velocities (see Figure~\ref{fig:velobs} and the left panel of Figure~\ref{fig:comparisons}). For this reason, we have used in our work only three free parameters, fixing the starting velocity of the cool CGM clouds at zero.
\end{appendix}

\end{document}